\documentclass[twoside,onecolumn,12pt]{article}

\usepackage{graphicx}
\usepackage{caption}

\usepackage[T1]{fontenc}                    % Use 8-bit encoding that has 256 glyphs
\usepackage[bitstream-charter]{mathdesign}
\usepackage[english]{babel}                 % Language hyphenation and typographical rules

\usepackage[a4paper,left=1.5cm,right=1.5cm,top=2.5cm,bottom=2.5cm]{geometry} % Document margins
\usepackage{lettrine}                       % The lettrine is the first enlarged letter at the beginning of the text

\usepackage{booktabs}                       % Horizontal rules in tables

\usepackage{titling}                        % Customizing the title section
\usepackage[noblocks]{authblk}
\usepackage{hyperref}                       % For hyperlinks in the PDF

\usepackage{url}

\usepackage{abstract}                       % Allows abstract customization
\newcommand{\upcite}[1]{\textsuperscript{\textsuperscript{\cite{#1}}}}
\newcommand{\upciteeg}[1]{\textsuperscript{\textsuperscript{\cite[for example]{#1}}}}

\newcommand{\Teff}{$T_{\rm eff}$}
\newcommand{\logg}{$\log g$}

\newcommand{\Vt}  {$\xi_t$}
\newcommand{\kms}{km\,s$^{-1}$}

\newcommand{\Eexc}{$E_{\rm exc}$}
\newcommand\ion[2]{#1$\;${\scriptsize\rmfamily\uppercase\expandafter{\romannumeral#2}}\relax}

\newcommand{\dpi}{$\Delta {\rm P}$}
\newcommand{\dnu}{$\Delta\nu$}
\newcommand{\numax}{$\nu_{\rm max}$}

\begin{document}
\pretitle{\begin{center}\Large} % Article title formatting
\posttitle{\end{center}} % Article title closing formatting

\title{\textbf{Most Lithium-rich Low-mass Evolved Stars Revealed as Red Clump stars by Asteroseismology and Spectroscopy}} % Article title
\author[1,$\ast$]{\small Hong-Liang Yan}
\author[1,12,13,$\ast$]{Yu-Tao Zhou}
\author[2]{Xianfei Zhang}
\author[2,14,15]{Yaguang Li}
\author[1,3]{Qi Gao}
\author[1,3, $\dagger$]{Jian-Rong Shi}
\author[1,3, $\dagger$]{Gang Zhao}
\author[4,5]{Wako Aoki}
\author[4,5,6]{Tadafumi Matsuno}
\author[7]{Yan Li}
\author[1]{Xiao-Dong Xu}
\author[1]{Haining Li}
\author[1]{Ya-Qian Wu}
\author[2]{Meng-Qi Jin}
\author[8]{Beno{\^i}t Mosser}
\author[2]{Shao-Lan Bi}
\author[2]{Jian-Ning Fu}
\author[9]{Kaike Pan}
\author[10,11]{Takuma Suda}
\author[1]{Yu-Juan Liu}
\author[1]{Jing-Kun Zhao}
\author[1,3]{Xi-Long Liang}

\affil[1]{\small CAS Key Laboratory of Optical Astronomy, National Astronomical Observatories, Beijing 100101, China.}
\affil[2]{Department of Astronomy, Beijing Normal University, Beijing 100875, P. R. China.}
\affil[3]{School of Astronomy and Space Science, University of Chinese Academy of Sciences, Beijing 100049, China.}
\affil[4]{National Astronomical Observatory of Japan, 2-21-1 Osawa, Mitaka, Tokyo 181-8588, Japan}
\affil[5]{Department of Astronomical Science, School of Physical Sciences, The Graduate University for Advanced Studies (SOKENDAI), 2-21-1 Osawa, Mitaka, Tokyo 181-8588, Japan.}
\affil[6]{Kapteyn Astronomical Institute, University of Groningen, Landleven 12, 9747 AD Groningen, The Netherlands}
\affil[7]{Yunnan Observatories, Chinese Academy of Sciences, Kunming 650011, Yunnan, China}
\affil[8]{LESIA, Observatoire de Paris, Universit\'e PSL, CNRS, Sorbonne Universit\'e, Universit\'e de Paris, 5 place Jules Janssen, 92195 Meudon, France}
\affil[9]{Apache Point Observatory and New Mexico State University, PO Box 59, Sunspot, NM88349-0059, USA.}
\affil[10]{The Open University of Japan, Wakaba 2-11, Mihama-ku, Chiba, 261-8586, Japan}
\affil[11]{Research Center for the Early Universe, University of Tokyo, Hongo 7-3-1, bunkyo-ku, Tokyo, 113-0033, Japan}
\affil[12]{Department of Astronomy, School of Physics, Peking University, Beijing 100871, People's Republic of China}
\affil[13]{Kavli Institute for Astronomy and Astrophysics, Peking University, Beijing 100871, People's Republic of China}
\affil[14]{Sydney Institute for Astronomy (SIfA), School of Physics, University of Sydney, NSW 2006, Australia}
\affil[15]{Stellar Astrophysics Centre, Department of Physics and Astronomy, Aarhus University, Ny Munkegade 120, 8000 Aarhus C, Denmark}
\affil[$\ast$]{These authors contribute equally to the paper}
\affil[$\dagger$]{corresponding authors: sjr@nao.cas.cn, gzhao@nao.cas.cn}

\date{} % Leave empty to omit a date

\maketitle

%----------------------------------------------------------------------------------------
%   ARTICLE CONTENTS
%----------------------------------------------------------------------------------------

%----------------------------------------------------------------------------------------
%   ABSTRACT
%----------------------------------------------------------------------------------------
\begin{center}
\textbf{Abstract}
\end{center}

Lithium has confused scientists for decades at almost each scale of the universe. 
Lithium-rich giants are peculiar stars with lithium abundances over model prediction. 
A large fraction of lithium-rich low-mass evolved stars are traditionally supposed to be red giant branch (RGB) stars. 
Recent studies, however, report that red clump (RC) stars are more frequent than RGB.
Here, we present a uniquely large systematic study combining the direct asteroseismic analysis with the spectroscopy on the lithium-rich stars.
The majority of lithium-rich stars are confirmed to be RCs, whereas RGBs are minor. 
We reveal that the distribution of lithium-rich RGBs steeply decline with the increasing lithium abundance, showing an upper limit around $2.6$ {\it dex}, whereas the Li abundances of RCs extend to much higher values. 
We also find that the distributions of mass and nitrogen abundance are notably different between RC and RGB stars.
These findings indicate that there is still unknown process that significantly affects surface chemical composition in low-mass stellar evolution.

%----------------------------------------------------------------------------------------
%   1. Brief Introduction \& Scope of the Study
%----------------------------------------------------------------------------------------
\begin{center}
\textbf{Introduction}
\end{center}

The chemical signatures of stars provide rich information of their origins and evolutions\upcite{Freeman2002, Spite1982}. The standard stellar evolution model predicts steep dilutions of lithium (Li) abundance in giant stars\upcite{Iben1967}, which have been confirmed by many observations\upcite{Brown1989, Lind2009}. However, a small fraction of giants are found to preserve anomalously high Li abundance\upcite{Wallerstein1982, Kumar2011, Martell2013, Smiljanic2018, Deepak2019, Casey2019, Gao2019}, some of them even exceeds model prediction by thousands of times\upcite{Yan2018}. These stars were called Li-rich giants, with a classic definition of $A_{\rm Li} \ge $ 1.5. Here, $A_{\rm Li}$ is the Li abundance expressed as $A_{\rm Li}= \log({\rm N_{Li}}/{\rm N_{H}}) + 12$, where ${\rm N_{Li}}$ and ${\rm N_{H}}$ is the number density of lithium and hydrogen, respectively. Li-rich giants evoke great interests as they reveal the origin and evolution of Li in stars. The Li-rich giants are traditionally supposed to be either red giant branch (RGB) stars or asymptotic giant branch (AGB) stars\upcite{Charbonnel2000} from their locations on the Hertzsprung-Russell (H-R) diagram, and theoretical interpretations also require stars to be in those phases\upcite{Sackmann1999, Denissenkov2004, Charbonnel2010}. However, the evolutionary phases of the RGB luminosity bump usually overlap with that of the red clump (RC) - a stable phase with core helium-burning after the helium flash - on the H-R diagram (the {\it overlapping region} hereafter). The first discovery of the Li-rich core helium-burning stars was based on the asteroseismic analysis\upcite{Silva2014}. With the growing number of such stars reported \upcite{Carlberg2015, Smiljanic2018, Kumar2018, Singh2019, Zhou2019, Casey2019, Martell2020}, Li-rich RC stars become crucially important. Very recently, Kumar et al. proposed that all the RC stars have enhanced Li abundance for their evolutionary stage\upcite{Kumar2020}, following the He-core flash at the RGB tip for which very low Li-abundances are predicted by standard stellar evolution models. The Li production in both RGB and RC stars are not well understood, and a series of major questions were raised: 1) how much fraction of Li-rich giants are RC stars and why; 2) what are the signatures of Li-rich RGB and RC stars, and 3) how was Li produced in these evolved stars?

%----------------------------------------------------------------------------------------
%   2. Results
%----------------------------------------------------------------------------------------
\begin{center}
\textbf{Results}
\end{center}

To address these important questions, here we report the largest systematic study combining the direct asteroseismic analysis with the spectroscopy on the Li-rich giants from common sources of the Large Sky Area Multi-Object Fiber Spectroscopy Telescope (LAMOST) survey\upcite{Cui2012} and Kepler\upcite{Borucki2010} Input Catalog (KIC). From the LAMOST low-resolution spectra, we obtained $\sim 42,000$ giant stars with surface gravity (\logg) below $3.5$. We derive the Li abundances of these giants using the template matching method based on the stellar parameters provided by LAMOST Data Release 7 (DR7). Though different definitions of Li-richness were proposed\upcite{Kumar2020, Kirby2016}, we adopt the most widely-used standard of $A_{\rm Li} = 1.5$ plus an uncertainty level of 0.2 to identify the Li-rich giants from our data. Finally, we found $455$ stars with $A_{\rm Li} \ge 1.7$ (the {\it low-resolution sample} hereafter). This sample covers a metallicity ([Fe/H]) range from -2.5 to +0.7 {\it dex}, though the majority of the stars have $-0.6 <$ [Fe/H] $< 0.4$. Similar to the previous study\upcite{Casey2019}, we found that the occurrence rate of Li-rich giants is higher around the solar metallicity. For comparison with the {\it low-resolution sample}, we also performed high-resolution spectroscopic observations to 26 stars in the {\it low-resolution sample} plus three stars from the Kepler `Second Light' (K2) mission (the {\it high-resolution sample} hereafter). Supplementary Table~\ref{sup_tab_1} shows the detailed observation information for the {\it high-resolution sample} stars. Their Li abundances are derived using the spectrum synthesis method from the $6707.8$\,\AA\ and $6103.6$\,\AA\ lines with the non-local thermodynamic equilibrium effect (NLTE) considered. 

To obtain their evolutionary phases from the Kepler asteroseismic data, we cross-matched our {\it low-resolution sample} stars with the classifications obtained from the method presented by Hon et al. (2017)\upcite{Hon2017} using the Kepler power spectrum. We found $134$ stars are in common with our {\it low-resolution sample}. Using their results as an initial input, we obtained the evolutionary phases from period spacings (\dpi), frequency separation (\dnu) and/or asteroseismic patterns\upcite{Hon2017, Bedding2011} for these stars. We also obtained their masses and radius from the scaling relations\upcite{Brown1991} based on the asteroseismic frequency separation and maximum frequency(\numax). We found that $115$ out of the $134$ Li-rich giants are in the core helium-burning phase, and the remaining $19$ stars are hydrogen-shell burning RGB stars.

\begin{figure}[!h]
\begin{center}
\includegraphics[angle=0, width=12cm]{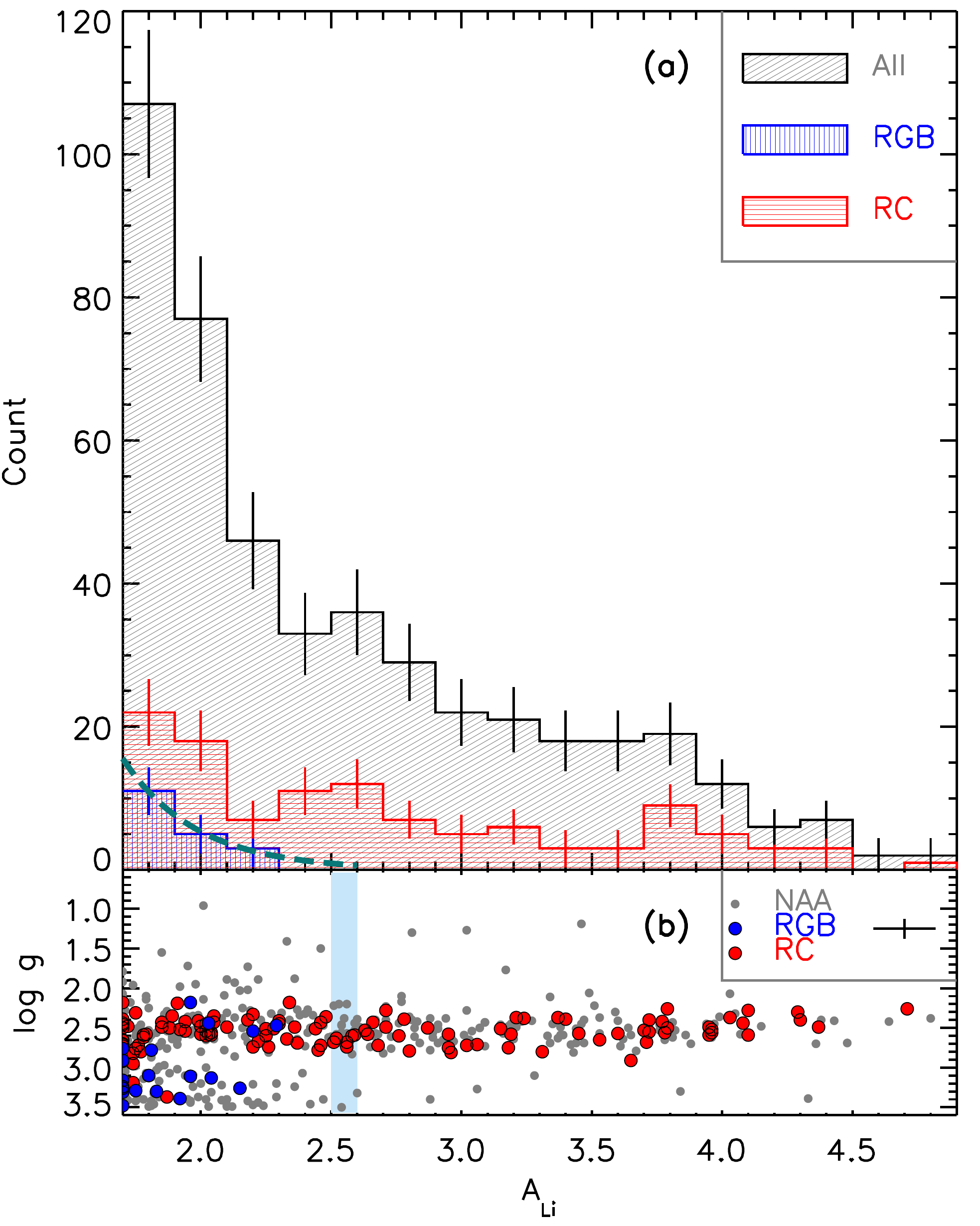}\\
\end{center}
\caption{\textbf{The distribution of Li abundance in the whole sample.} In the top panel, the bin size for Li abundance is 0.20 {\it dex}. Stars with evolutionary phases derived from asteroseismology are indicated with red (for RC stars) and blue (for RGB stars). The whole sample of $455$ Li-rich giants are indicated with color grey. The dashed line in olive green represents the exponential function fitting to the distribution function of Li-rich RGB giants and has been scaled to match the vertical axis scale. The bottom panel shows the Li abundances of each star as a function of \logg. The symbol colors are similar to the top panel. `NAA' means there is `no asteroseismic analysis'. In the top panel, error bar represents $1\sigma$ statistical uncertainty of the number of stars per bin. Typical errors of \logg\ and Li abundance are shown in the legend of the bottom panel, which are 0.25 dex and 0.24 dex, respectively (see `Methods' for error estimation). The upper limit of the Li abundance for RGB stars is highlighted in light blue. The upper abundance limit is defined at the frequency of $\displaystyle e^{-3}$, which corresponds to $A_{\rm Li} \simeq 2.6\pm0.24$ dex.}\label{fig_1}
\end{figure}

In Fig.~\ref{fig_1}, we show the distribution of the Li abundance for the whole sample, with RC and RGB stars identified from the asteroseismic analysis color-coded. Here we define the \emph{ratio} of Li-rich RC stars as $r_{\rm{RC}}(A_{\rm Li}^{'})=N_{\rm{RC}}/(N_{\rm{RC}}+N_{\rm{RGB}}) \times 100\,\% $, where $N_{\rm{RC}}$ and $N_{\rm{RGB}}$ represent the number of Li-rich RC stars and Li-rich RGB stars in a certain Li abundance range $A_{\rm Li}^{'}$, respectively. Considering the whole sample of stars with asteroseismic evolutionary phases, we found that the overall $r_{\rm{RC}}$ is $86\,\%$, which suggests that the majority of Li-rich giants are core helium-burning RC stars. This result is consistent with some of the most recent studies\upcite{Casey2019}, but importantly, it is obtained from the direct use of asteroseismic information, thus the evolutionary phase of Li-rich giants is more robust. We found that the frequency of Li-rich RC stars through the entire RC sample of Hon's\upcite{Hon2017} is $1.5\%$, while it is only $0.25\%$ for the Li-rich RGB stars. The high frequency of Li-rich stars at the RC stage is essentially important, as it provides the most significant constraint on the Li production. Many classic theories\upcite{Sackmann1999, Denissenkov2004, Charbonnel2010} accounting for Li enhancing are only suitable for interpreting Li-rich phenomenon in RGB stars, rather than RC stars. 

\begin{figure}[!h]
\begin{center}
\includegraphics[angle=0, width=\hsize]{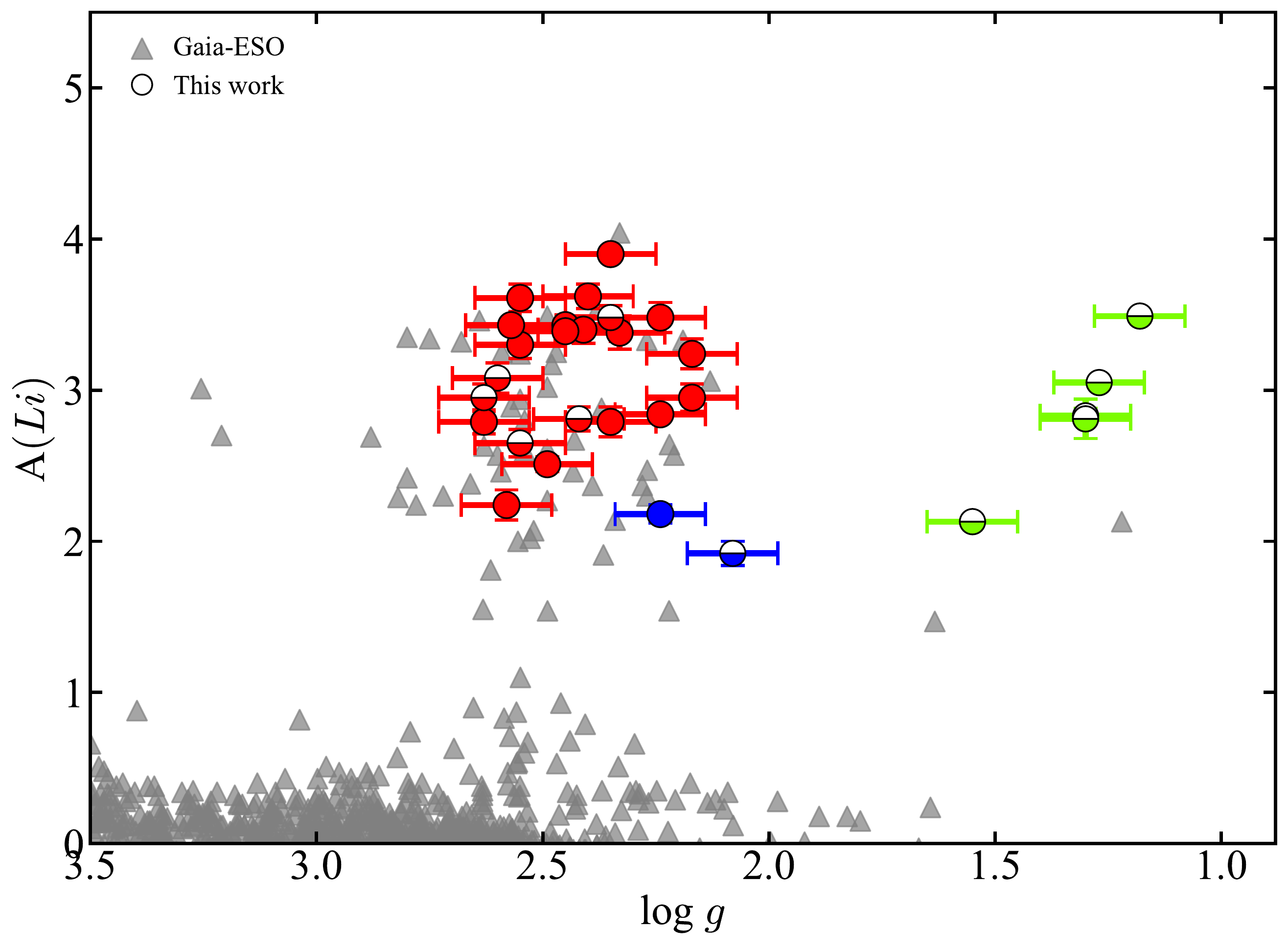}\\
\end{center}
\caption{\textbf{The Li abundance and classification of the high-resolution sample.} The abundance is plotted as a function of \logg\ for our high-resolution spectroscopic targets (colored dots) with the Li-rich giants discovered by the Gaia-ESO DR3 (triangles). For our {\it high-resolution sample}, dots with full-filled colors represent stars with evolutionary phase derived from the asteroseismic data, while dots with half-fulled colors represent stars with evolutionary phase derived from the H-R digram (see `Methods' section). The RGB and RC stars are indicated with blue and red, respectively. Green symbols are highly evolved RGB or AGB stars. The error bars of \logg\ and Li abundance are shown in the figure for each star (see `Methods' for error estimations of the \logg\ and luminocity.)}\label{fig_2}
\end{figure}

The Li abundances derived from our {\it low-resolution sample} agree with those from the high-resolution data based on the resonance and subordinate lines (see `Methods' for details). Moreover, the high frequency of RC stars in the {\it overlapping region} is confirmed in the {\it high-resolution sample} as well. We obtained the evolutionary phases for 18 Li-rich giants with asteroseismic data in the {\it high-resolution sample}, and found that the RC stars is also dominant (Fig.~\ref{fig_2}). In the {\it overlapping region}, the $r_{\rm{RC}}$ obtained from the {\it high-resolution sample} is higher than that of the {\it low-resolution sample}, which is due to the sample selection that stronger Li lines (higher Li abundance) are more preferred during the selection of targets for high-resolution spectroscopy. The stellar parameters and Li abundances of our {\it high-resolution sample} stars are listed in Supplementary Table~\ref{sup_tab_2} (see `Methods' for details).

Furthermore, we found that the distribution of Li-rich RC stars in the {\it low-resolution sample} is significantly different from that of the RGB stars. The distribution of Li-rich RC stars in Fig.~\ref{fig_1} covers a wide range of $A_{\rm Li}$ from $1.7$ {\it dex} to $4.8$ {\it dex}, and they are more Li-rich than the RGB stars on avenge. For the Li-rich RGB stars, the distribution shows significantly steep decline with increasing $A_{\rm Li}$. In addition, there seems an upper abundance limit in the distribution of Li-rich RGB stars in our sample, as shown in Fig.~\ref{fig_1} (both top and bottom panels). We found no RGB star with Li abundance over $2.4$ {\it dex} in our sample, which implies that there might be an upper limit of the Li abundance in RGB stars. If so, the upper limit can be estimated by fitting the distribution with a decreasing function. The best fit to the distribution is expressed as {$\displaystyle y=e^{(-3.6x_{i}+6.4)}\times 100\,\%$}, where $x_{i}$ represents the average Li abundance in the $i$th bin, and $y$ is the ratio of the number of stars in the $i$th bin to that in the first bin. Here, the first bin consists of stars with Li abundance between 1.7 and 1.9 {\it dex}, and is defined as normalization. The upper limit of Li abundance can be estimated by setting $\displaystyle y=e^{-3}$, which means that the frequency of RGB stars with $A_{\rm Li}>2.6$ becomes lower than $e^{-3}$ than those with $A_{\rm Li}=1.7$. Considering the error of the Li abundance derived from our {\it low-resolution sample} (see `Methods' for details), we estimated the uncertainty of the upper abundance limit as $\pm 0.24$ {\it dex}. 

Owing to the low frequency of Li-rich RGB stars, the sample size is too small for a more sophisticated statistic analysis, although our sample is relatively large in the Kepler field with available asteroseismic data. Nevertheless, the distribution of Li abundance for {\it all} the stars with asteroseismology data is representative of the distribution for the whole sample based on the result of Kolmogorov-Smirnov (K-S) test. Their significance level of having the same distribution is 0.94, and their maximum deviation is 0.05. This indicates that the steep decline of Li-rich RGB stars towards higher Li abundance is applicable through the whole sample. Our Li abundance is derived from the template matching method, which gives more reliable results at higher Li abundance as the matching becomes easier with the Li line growing stronger. This means that the possibility of missing-out stars towards the higher Li abundance is very low. Furthermore, previous studies\upciteeg{Singh2019} from which the evolutionary phases of Li-rich giants were determined by the direct asteroseismic analysis found no Li-rich RGB stars beyond the upper limit either. On the other hand, however, there could be selection bias due to the asteroseismology. Since the stellar activity can damp the asteroseismic signal\upcite{Chaplin2011}, asteroseismology is not available for stars with strong activity. We may miss such stars if the Li-excess is strongly related to the stellar activity. These stars are expected in objects with low surface gravity. Indeed, there are about 20 Li-rich stars with \logg $< 2.0$ for which asteroseismology data are not available at present. They would be highly evolved RGB or AGB stars. Among them, only four stars have A(Li)>2.6. This indicates that, even if they are RGB stars rather than AGB, we conclude that the Li abundance distribution of Li-rich RGB stars is clearly different from that of RC stars. It is desired to confirm this result by more stars with asteroseismic data from, for example, Transiting Exoplanet Survey Satellite (TESS)\upcite{Ricker2015}. We note that the three stars with $A_{\rm Li}>2.6$ and \logg $< 2.0$ in the {\it high-resolution sample} overlap the four such objects in the {\it low-resolution sample} mentioned above. This clearly indicates the high fraction of such objects in the high-resolution sample is due to a bias in the sample selection.

\begin{figure}[h!]
\begin{center}
\includegraphics[angle=0, width=10cm]{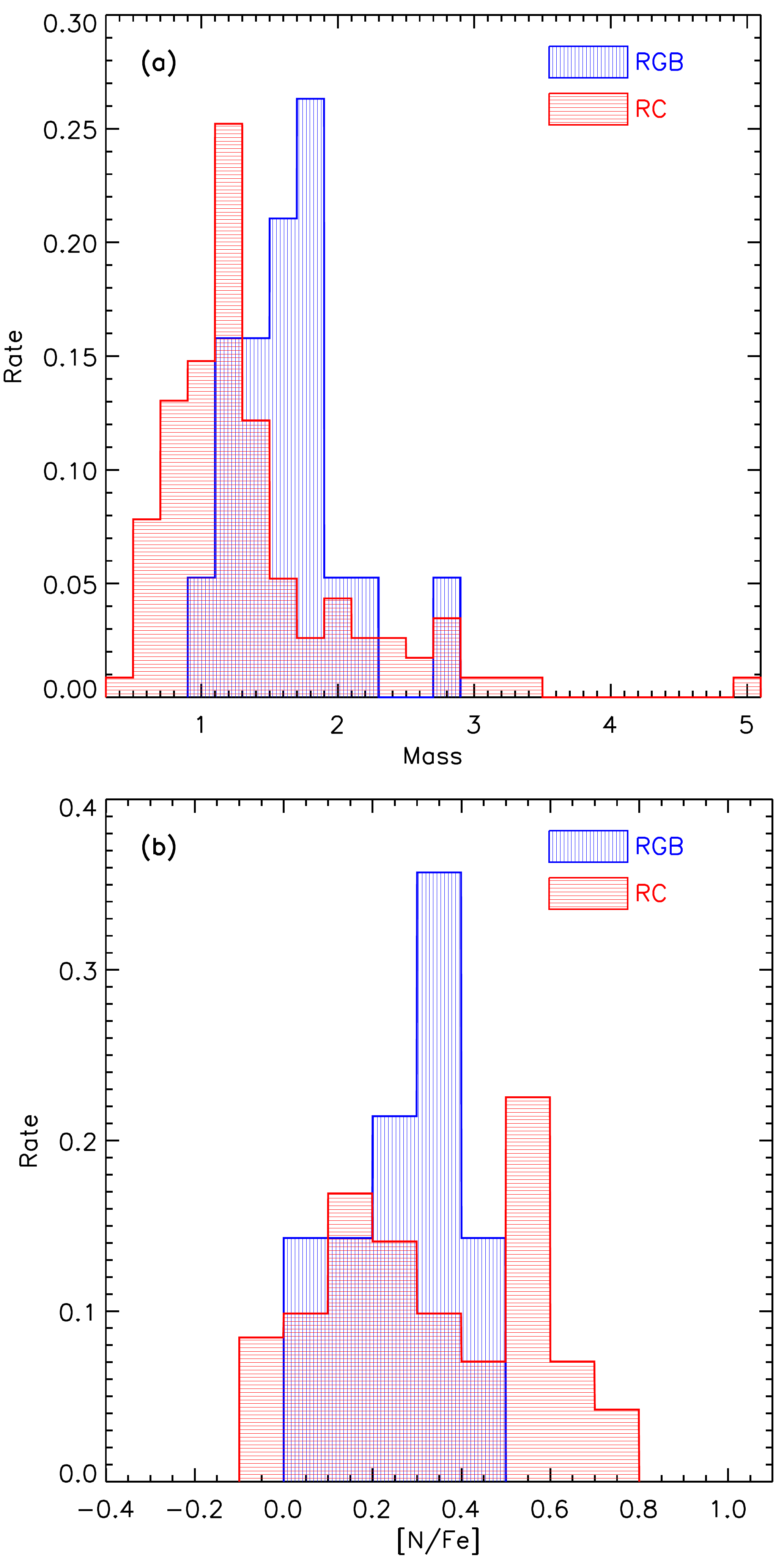}\\
\end{center}
\caption{\textbf{The distribution of mass and [N/Fe] in the sample.} The distribution is plotted as a function of mass (top panel) and [N/Fe] (bottom panel). The bin size for mass and [N/Fe] are 0.2 M$_{\odot}$ and 0.1 {\it dex}, respectively. Similar to other figures, stars with asteroseismic evolutionary phases are indicated with red (for RC stars) and blue (for RGB stars).}\label{fig_3}
\end{figure}

Finally, we found that the Li-rich RGB and RC stars have different distribution in mass and nitrogen (N) abundance, as shown in Fig.~\ref{fig_3}. The mass is calculated from the scaling relations and the N abundance is adopted from APOGEE Stellar Parameters and Chemical Abundances Pipeline (ASPCAP)\upcite{Garcia2016} Data Release 15 (DR15). Interestingly, the Li-rich RGB stars seem to have a larger mass on average than the Li-rich RC stars. The mass distribution peaks at 1.7 M$_{\odot}$, which is also higher than that of Li-normal stars (Extended Data Fig.~\ref{ext_fig_1}, top-left panel). The engulfment scenario seems to coincide with this signature. The distribution of [N/Fe] is also striking, as the Li-rich RGB stars show a single peak at 0.4, while the Li-rich RC stars seem to have double peaks. The second [N/Fe] peak for Li-rich RC stars covers the range from $\sim 0.40$ to $\sim 0.80$ {\it dex}, with a maximum distribution at $\sim 0.60$ {\it dex}. What is at odd is that the second [N/Fe] peak is not seen for the RC stars without Li enhancing, as shown in Extended Data Fig.~\ref{ext_fig_1}. Further work will be needed to explain this difference.

%----------------------------------------------------------------------------------------
%   3. Discussions
%----------------------------------------------------------------------------------------
\begin{center}
\textbf{Discussions}
\end{center}

\noindent\textbf{Possible Interpretations to the Signatures of Li-rich RGB Stars.} 

The Li-rich giants have been found for decades, and a number of scenarios\upcite{Alexander1967, Cameron1971, Sackmann1999, Denissenkov2004, Charbonnel2010, Lebzelter2012, Kirby2016, Yan2018, Martell2020, Kumar2020, Charbonnel2020} were proposed to interpret their origins. The steep decline of the distribution and the upper abundance limit in Li-rich RGB stars have now been revealed. It indicates that the Li-rich RGB stars can only be interpreted by the scenarios with Li abundances not higher than $2.6$ {\it dex}. Given this constraint, two possible scenarios still work. The first one is that Li-rich RGB stars are only a natural consequence of Li depletion by the first dredge-up (FDU) process. Assuming that the abundance in Li is diluted by a factor of $\sim 60$ by FDU\upcite{Iben1967}, the logarithmic abundance will decrease by $\sim 1.8$ {\it dex}. An RGB star with Li abundance at $2.6$ {\it dex} is estimated to have Li abundance of $4.4$ {\it dex} before FDU, for example, at the turn-off stage. Such Li-rich predecessor had indeed been reported for metal-poor stars\upcite{Li2018}. The second one is the engulfment of a giant planet\upcite{Aguilera2016}. Unlike the internal production of Li, the engulfment scenario was usually thought to be unable to account for the super Li-rich objects. If the upper abundance limit is true, this scenario would be possible, although the upper limit reported here is slightly higher than the predicted maximum Li enhanced by engulfment event. Fig.\ref{fig_1} also shows the lack of Li-rich bright giants with \logg $< 2.2$. This coincides with the depletion of Li and carbon at the RGB bump in metal-poor giants\upcite{Gratton2000}, probably caused by extra mixing due to the disappearance of the barrier of the mean molecular weight around the hydrogen burning shell\upcite{Charbonnel2010}.\\

\noindent\textbf{Possible Interpretations to the Signatures of Li-rich RC Stars.}

In recent years, the use of asteroseismic data based on the time-domain photometry obtained from space telescope such as Kepler\upcite{Borucki2010}, CoRoT\upcite{Auvergne2009} and TESS\upcite{Ricker2015} resulted in a growing number of Li-rich RC stars\upcite{Silva2014, Carlberg2015, Smiljanic2018, Kumar2018, Singh2019, Casey2019}. Moreover, a large amount of such stars were obtained from machine learning although they might not have direct asteroseismic data\upcite{Casey2019, Zhou2019, Kumar2020}. Based on these observations, the ratio of RC stars to RGB stars among Li-rich giants is predicted to be $\sim$40-100\,\%. The wide range is mainly caused by the  sample size and the threshold used to define Li-richness\upcite{Casey2019, Singh2019, Zhou2019}. In this work, we showed that the ratio varies with $A_{\rm Li}$, and the integrated $r_{\rm{RC}}$ for the whole range of $A_{\rm Li}$ is $86\,\%$ based on the threshold of $A_{\rm Li}=1.7$. If the threshold is defined as $A_{\rm Li}=1.5$, then the $r_{\rm{RC}}$ will slightly decrease to $\sim 75\,\%$ in our sample. The result is consistent with the ratio presented by the recent work\upcite{Casey2019}. While for the stars with $A_{\rm Li} \ge 2.6$ {\it dex}, $r_{\rm{RC}}$ can reach $100\,\%$\upcite{Singh2019}. We note that these numbers significantly change if we adopt different criteria for RC stars from RGB stars, as the recent study of Kumar et al. proposes that the standard for Li-richness of RC stars should be defined separately as $A_{\rm Li} \ge -0.9$\upcite{Kumar2020}.

The presence of Li-rich RC stars impacts the previous understanding of the Li production. Only a few scenarios are proposed to interpret the Li-enhancing in RC stars. This is partly because only until recently that RC stars are recognized as the dominant population\upcite{Singh2019,Casey2019,Kumar2020} among Li-rich evolved stars. The difficulty of the Li production in RC stars might be another reason. As shown in Fig.~\ref{fig_1}, the RC stars are more Li-rich than the RGB stars on avenge. This means that the strong Li-enhancement might occur during the helium core-flash or at the RC phase. If so, there is a higher probability of the Li-enhancement at the tip of the RGB due to the presence of the convective envelope by which nuclear products inside could be dredged-up. Kumar et al. also propose that Li is produced between the stages of RGB-tip and RC, suggesting that all the low-mass stars undergo this production\upcite{Kumar2020}. We cannot identify satisfactory models or scenarios to enhance Li during the core-helium flash yet. The hydrogen entrainment by the helium flash convective zone for low-metallicity AGB stars may give us a hint\upcite{Iwamoto2004}. However, the nucleosynthesis result for the hydrogen mixing during the helium-core flash for a $1$~M$_{\odot}$ model with [Fe/H] $= -6.5$ does not favor a strong Li enhancement\upcite{Campbell2010}. 

\begin{figure}[!h]
\begin{center}
\includegraphics[angle=0, width=\hsize]{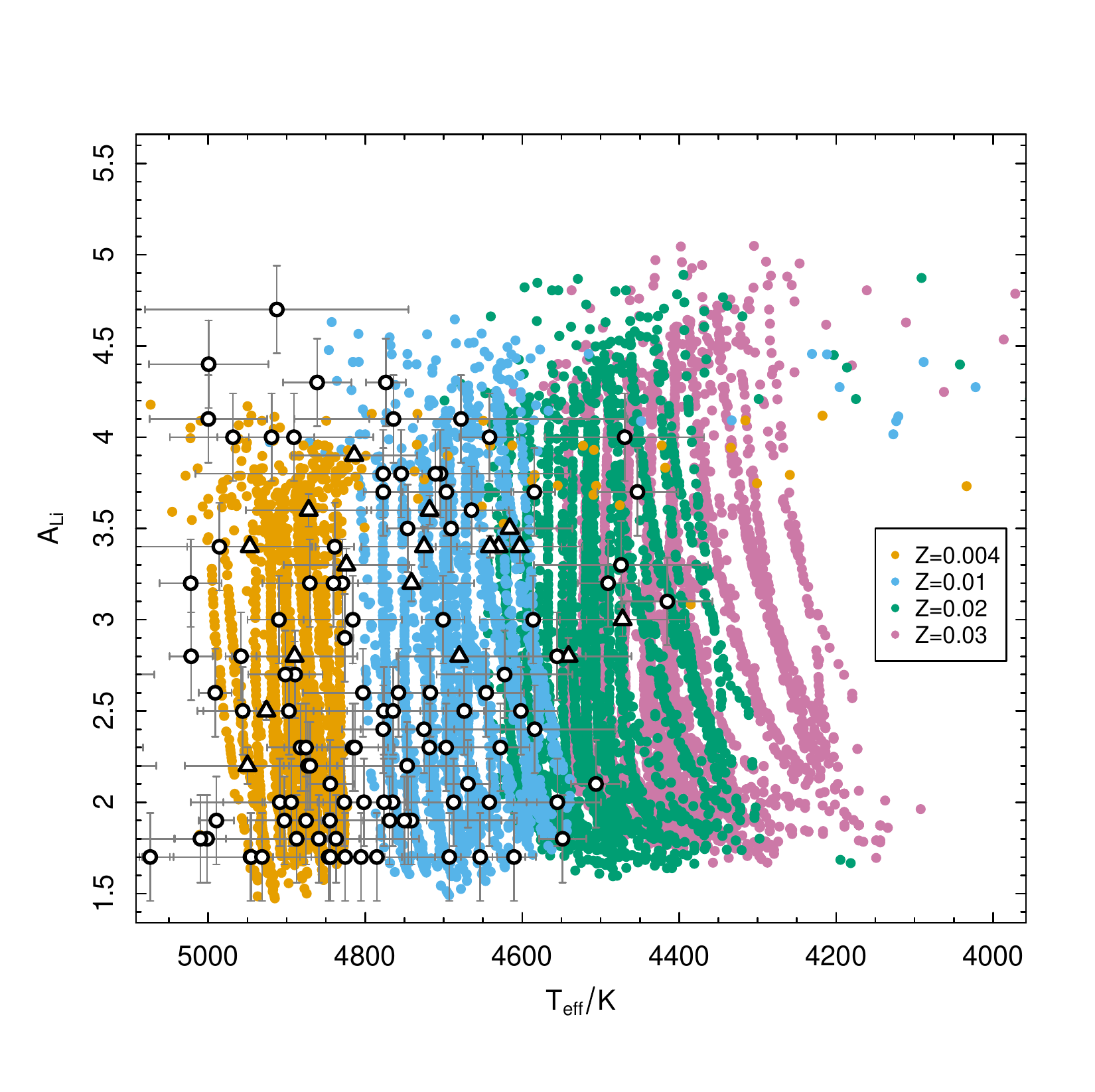}\\
\end{center}
\caption{\textbf{Comparison of the Li abundances between the HeWD-RGB merger model prediction and the stars in our sample.} The predicted Li abundances in different metallicities are indicated with different colors. The metallicity range is as same as that in Fig.~\ref{fig_2} but only with a larger step. The scatter of the Li abundance in each metallicity is caused by the different mass ratio of the merged HeWD and RGB star. RC stars from the {\it high-resolution sample} are indicated with triangles, while RC stars from the {\it low-resolution sample} are indicated with circles. The error bars of \Teff\ and Li abundance are shown for each star in the figure with grey solid lines (see `Methods' for error estimations of the \Teff\ and Li abundance).}\label{fig_4}
\end{figure}

Zhang \& Jeffery proposed\upcite{Zhang2013} and improved\upcite{Zhang2020} a scenario that the merger of a helium white dwarf (HeWD) with a RGB star could trigger a convection shell which synthesizes Li and leads to a Li-rich RC star. Based on this scenario, we calculated the Li abundance from a series of improved merger models and combined it with the binary population synthesis (see `Methods' for details). In Fig.~\ref{fig_4}, we show our Li abundances derived from the high- and low-resolution spectra in a $A_{\rm Li}$ - \Teff\ plane along with the predicted Li abundances by the model as the background. It seems that the HeWD-RGB merger scenario generally agrees with the observed Li abundances in our sample. Furthermore, the HeWD-RGB merger model predicts that most of the resulted RC stars have masses between $0.8-1.8$ M$_{\odot}$ with a peak around $1.1-1.2$ M$_{\odot}$, which is generally consistent with our result (Fig.~\ref{fig_3}) except for a tail in the high mass end. The model also predicts enhancement of N abundance. Indeed, a fraction of stars show larger [N/Fe] in our sample (Fig.~\ref{fig_3}), but there seems no evident correlation\upcite{Martig2016} between Li-rich stars with high mass and high [N/Fe] (Extended Data Fig.~\ref{ext_fig_2}). We find that the HeWD-RGB merger model can explain a set of signatures observed in Li-rich RC stars except for the N abundance.

The non-conventional mixing could result in a change of N abundance\upcite{Charbonnel2010, Lagarde2012} for RGB stars. However, there seems very few discussion about non-conventional mixings in RC stars. Meanwhile, the metallicity could affect initial chemical composition including N. But for stars that have evolved to RGB or later stages, their N abundances are also affected by the convection processes, such as the dredge-up and the extra mixing\upcite{Lagarde2012}. Thus, the role of metallicity becomes less important than that played in the main sequence phase. Metallicity could also play a role in Li-production. For example, if the Li is produced by planet engulfment or the binary interaction, there would be more Li-rich giants at the higher metallicity as planets and binaries also favor the high metallicity environment\upcite{Gonzalez2014}. But if the Li is from novae pollution, one would expect a lower rate of Li-rich giants, as the explosion rate of nova should be lower in the environment of higher metallicity\upcite{Gao2017,Grisoni2019}.

\begin{center}
\textbf{Summary}
\end{center}

In this work, we present a systematic study combining the asteroseismology with the spectroscopy on the low-mass Li-rich evolved stars from the Kepler field. We reveal the steep decline of the distribution and the possible upper limit of Li abundance for RGB stars, wheres the RC stars are more frequently Li-rich than RGB stars. The distributions of mass and [N/Fe] are different between RC and RGB stars. We confirm that the majority of Li-rich evolved stars are at core helium-burning stage from the direct asteroseismic analysis. Although this result is derived statistically by recent studies without seismology, the use of seismic data enables us to investigate distributions of Li, N, and mass distributions (and more detailed properties in future studies) by classifying individual stars into RC and RGB. These findings provide new insights into the formation of Li-rich evolved stars. Further studies, from both observational and theoretical perspective, are urgently needed to interpret the Li production in the evolved stars.

%----------------------------------------------------------------------------------------
%   Main References
%----------------------------------------------------------------------------------------

\noindent \textbf{Correspondence and requests for materials should be addressed to}: J.-R.S. (sjr@nao.cas.cn) or G.Z. (gzhao@nao.cas.cn)\\

%----------------------------------------------------------------------------------------
%   Acknowledgements
%----------------------------------------------------------------------------------------
\small
\noindent \textbf{Acknowledgements} This research is supported by National Key R\&D Program of China No.2019YFA0405502, National Natural Science Foundation of China under grant Nos. 11988101, 11833006, 11833002, 11890694, 11973052, 11973042, 11603037, 11973049, and Strategic Priority Research Program of Chinese Academy of Sciences, Grant No. XDB34020205. We acknowledges the support from international partnership program's Key foreign cooperation project (No. 114A32KYSB20160049), Bureau of International Cooperation, Chinese Academy of Sciences. H.-L.Y. the supports from Youth Innovation Promotion Association (No. 2019060), Chinese Academy of Sciences. T.M. is supported by Grant-in-Aid for JSPS Fellows (grant No. 18J11326). K.P. acknowledges supports from the Mt. Cuba Astronomical Foundation Grant. This work is partially based on data collected at the Subaru Telescope, which is operated by the National Astronomical Observatory of Japan, and was supported by JSPS - CAS Joint Research Program. We acknowledge the support of the staff of the Lijiang 2.4m and 1.8m telescopes, and the support of Telescope Access Program (TAP) for accessing APF telescope. We acknowledge the supports from The LAMOST FELLOWSHIP that is supported by Special Funding for Advanced Users, budgeted and administrated by Center for Astronomical Mega-Science, Chinese Academy of Sciences (CAMS), and the supports from the Astronomical Big Data Joint Research Center, co-founded by the National Astronomical Observatories, Chinese Academy of Sciences and the Alibaba Cloud. Guoshoujing Telescope (LAMOST) is a National Major Scientific Project built by the Chinese Academy of Sciences. Funding for the project has been provided by the National Development and Reform Commission. LAMOST is operated and managed by the National Astronomical Observatories, Chinese Academy of Sciences.  We acknowledge the use of Gaia, APOGEE and Gaia-ESO data, and of VizieR catalogue access tool. We also thank J.-J. Mao and X.-T. Fu for discussions.

%----------------------------------------------------------------------------------------
%   Author Contributions
%----------------------------------------------------------------------------------------
\vspace{10pt}
\noindent \textbf{Author contributions} H.-L.Y., J.-R.S. and G.Z. proposed and designed this study. H.-L.Y, Y.-T.Z. and J.-R.S. led the data analysis with the contributions from Q.G., W.A., T.M., X.-D.X., H.L., and Y.-J.L.. X.Z., Yan Li., S.-L.B., and T.S. contributed to the model calculations and discussions. Yaguang Li., Y.-Q.W., M.-Q.J., B.M., and J.-N.F performed the asteroseismology analysis and derived the evolutionary phases. W.A., H.L., and K.P. carried out the high-resolution spectroscopic observations. J.-K.Z. and X.-L.L. performed the statistical calculation and tests. All the authors discussed the results and contributed to the writing of the manuscript.

%----------------------------------------------------------------------------------------
%   Author Information
%----------------------------------------------------------------------------------------
\vspace{10pt}
\noindent \textbf{Author Information} {
H.-L.Y. and Y.-T.Z. contribute equally to the paper. The authors declare no competing financial interests.
}

%----------------------------------------------------------------------------------------
%   Display Items
%----------------------------------------------------------------------------------------
\clearpage
%\noindent \textbf{Main Figure Legends}
\captionsetup[figure]{labelfont={bf},name={Fig.},labelsep=colon}

%%%%%%%%%%%%%%%%%%%%%%%%%%%%%%%%%%%%  Figure 1  %%%%%%%%%%%%%%%%%%%%%%%%%%%%%%%%%%%% 
%%%%%%%%%%%%%%%%%%%%%%%%%%%%%%%%%%%%  Figure 1  %%%%%%%%%%%%%%%%%%%%%%%%%%%%%%%%%%%% 

%%%%%%%%%%%%%%%%%%%%%%%%%%%%%%%%%%%%  Figure 2  %%%%%%%%%%%%%%%%%%%%%%%%%%%%%%%%%%%% 
%%%%%%%%%%%%%%%%%%%%%%%%%%%%%%%%%%%%  Figure 2  %%%%%%%%%%%%%%%%%%%%%%%%%%%%%%%%%%%% 

%%%%%%%%%%%%%%%%%%%%%%%%%%%%%%%%%%%%  Figure 3  %%%%%%%%%%%%%%%%%%%%%%%%%%%%%%%%%%%% 
%%%%%%%%%%%%%%%%%%%%%%%%%%%%%%%%%%%%  Figure 3  %%%%%%%%%%%%%%%%%%%%%%%%%%%%%%%%%%%% 

%%%%%%%%%%%%%%%%%%%%%%%%%%%%%%%%%%%%  Figure 4  %%%%%%%%%%%%%%%%%%%%%%%%%%%%%%%%%%%% 
%%%%%%%%%%%%%%%%%%%%%%%%%%%%%%%%%%%%  Figure 4  %%%%%%%%%%%%%%%%%%%%%%%%%%%%%%%%%%%% 

%----------------------------------------------------------------------------------------
%   Extended Data Figure
%----------------------------------------------------------------------------------------
\clearpage
\setcounter{figure}{0}
\captionsetup[figure]{labelfont={bf},name={Extended Data Fig.},labelsep=period}

%%%%%%%%%%%%%%%%%%%%%%%%%%%%%%%%%%%%  Extended Data Figure 1  %%%%%%%%%%%%%%%%%%%%%%%%%%%%%%%%%%%% 
\begin{figure}[!h]
\begin{center}
\includegraphics[angle=0, width=\hsize]{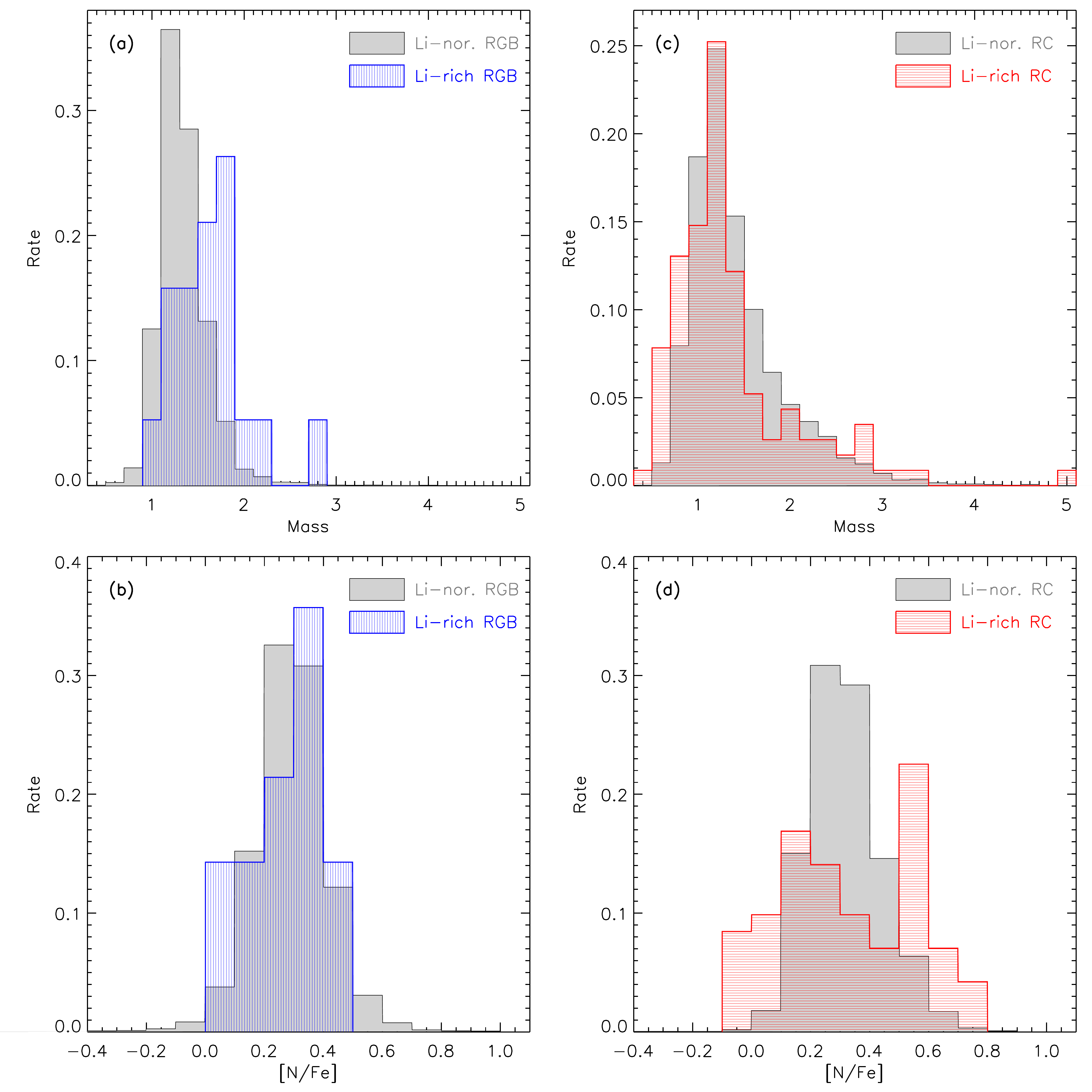}\\
\end{center}
\caption{\textbf{The distributions of mass and [N/Fe] for the Li-rich stars and Li-normal stars.} The gray histogram indicate the Li-normal RGB stars or RC star (as noted in each panel) of which the evolutionary phases identified by Hon et al. 2017\upcite{Hon2017}, the masses are from Yu et al. 2018\upcite{Yu2018}, and the [N/Fe] are from ASPCAP\upcite{Garcia2016}. The bin size is as same as Fig.~\ref{fig_3}}\label{ext_fig_1}
\end{figure} 
%%%%%%%%%%%%%%%%%%%%%%%%%%%%%%%%%%%%  Extended Data Figure 1  %%%%%%%%%%%%%%%%%%%%%%%%%%%%%%%%%%%% 

%%%%%%%%%%%%%%%%%%%%%%%%%%%%%%%%%%%%  Extended Data Figure 2  %%%%%%%%%%%%%%%%%%%%%%%%%%%%%%%%%%%% 
\begin{figure}[!h]
\begin{center}
\includegraphics[angle=0, width=\hsize]{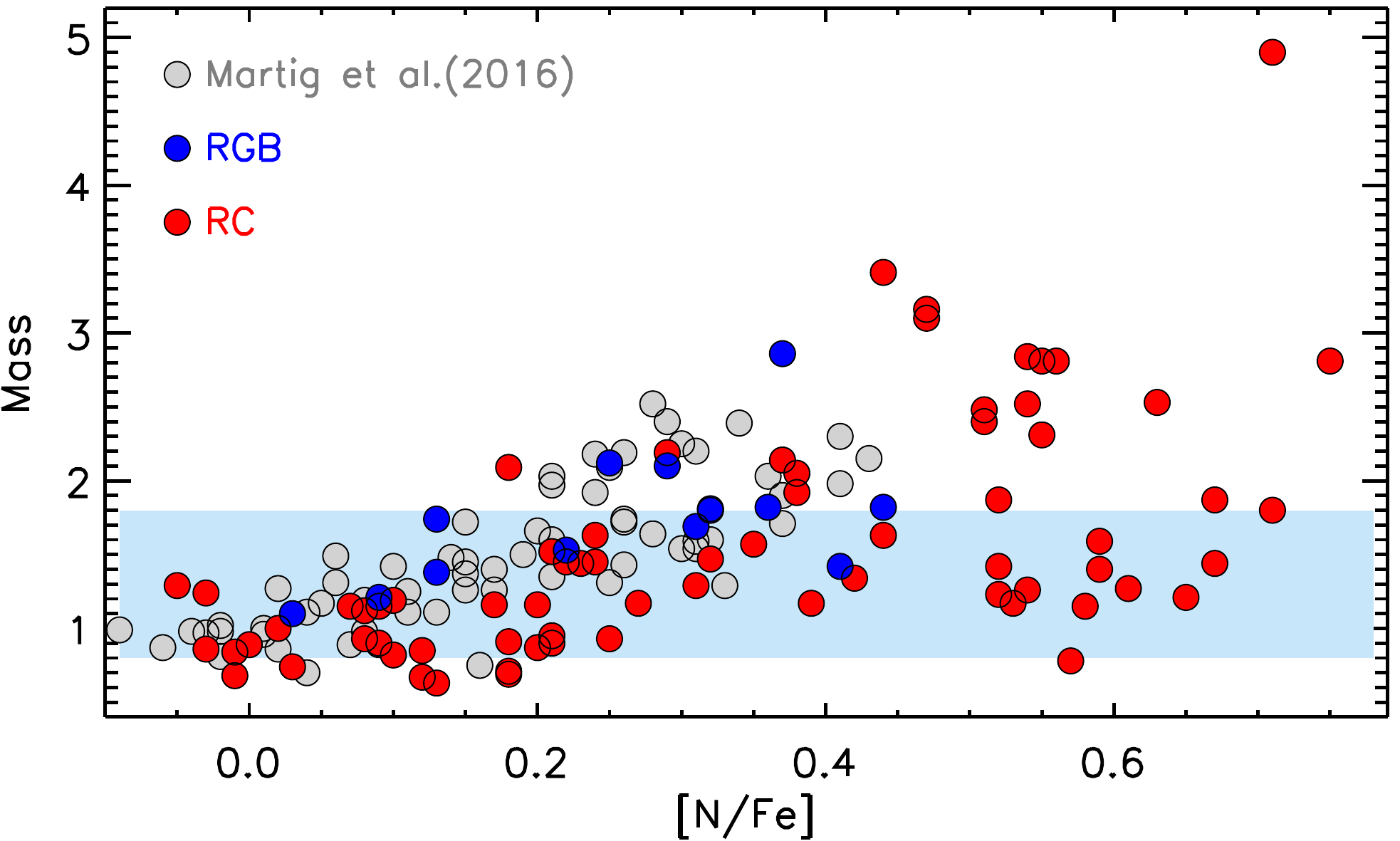}\\
\end{center}
\caption{\textbf{The mass versus [N/Fe] in our sample.} Martig et al. (2016) mapped the correlations\upcite{Martig2016} of stellar masses with ASPCAP [N/Fe] using machine learning. Their `predicted' masses of the common sources in our sample are plotted as the grey dots. The red and blue dots are Li-rich RC and RGB stars in our sample, respectively. The region marked with light blue is the most clumped mass range predicted by the HeWD-RGB merger model. For the predicted mass range of 0.8-1.8 M$_{\odot}$ by the HeWD-RGB model, a fraction of stars indeed show much larger [N/Fe] (also see Fig.~\ref{fig_3}) compared to other RC stars within this mass range, but there seems no evident correlation between stars with high-mass and stars with high-[N/Fe].}\label{ext_fig_2}
\end{figure} 
%%%%%%%%%%%%%%%%%%%%%%%%%%%%%%%%%%%%  Extended Data Figure 2  %%%%%%%%%%%%%%%%%%%%%%%%%%%%%%%%%%%% 

%----------------------------------------------------------------------------------------
%   Methods
%----------------------------------------------------------------------------------------
\clearpage
\appendix
\section*{Methods}

%----------------------------------------------------------------------------------------
%   Sample Selection
%----------------------------------------------------------------------------------------
\noindent\textbf{Sample Selection} 

The {\it low-resolution sample} is selected from a comprehensive sample of common sources of LAMOST and KIC. The total amount of common giant stars between this two datasets is $\sim 42,000$. Among those common sources, about $80\,\%$ are observed in an intense program named as `LAMOST-Kepler project', which aims to systematically survey for about $200,000$ stars with Kepler photometry\upcite{DeCat2015, Zong2018}. The other $20\,\%$ sources are observed in the regular survey. We first pick out all the common sources with \Teff\ $< 5,600$\,K and \logg $< 3.5$ to form a catalog of giant stars (here and after, the {\it giant catalog}), and then derive the Li abundances by template matching method. Considering an random error of $\sim 0.2$ {\it dex} of the method, we select stars with $A_{\rm Li} \ge 1.7$ in the {\it giant catalog} into our sample. This results in 455 Li-rich giants (referred as the {\it low-resolution sample}). The stellar parameters are obtained from LAMOST Stellar Parameter pipeline (LASP)\upcite{Luo2015}.

For the Li-rich giants of our {\it low-resolution sample}, the stars with sufficient asteroseismic data for identifying their evolutionary phases are selected by cross-matching the {\it low-resolution sample} with the classification obtained by Hon et al. (2017)\upcite{Hon2017}, which is based on the data of Kepler power spectra. Their classification covers $\sim 16,000$ stars, of which $7703$ stars are classified as RC stars and $7685$ stars are classified as RGB stars. Among them, we found 115 Li-rich stars are in RC phase, and the rest 19 stars are in RGB phase. 

The stars in our {\it high-resolution sample} are selected from Li-rich {\it candidates} obtained from the same {\it giant catalog}. These {\it candidates} are obtained by measuring the equivalent width of \ion{Li}{1} line at $6707.8$\,\AA, which is a separate procedure from the template matching. This means that the {\it candidates} with the high-resolution spectroscopic observations are selected before measuring their Li abundances and asteroseismic features. Among these {\it candidates}, 26 stars are observed with the high-resolution spectroscopy, and all of them are confirmed to be Li-rich giants. In addition, three stars from Kepler `Second Light' (K2) mission are added to be observed by the high-resolution spectroscopy in hoping of obtaining their evolutionary phases from K2 data. Finally, our {\it high-resolution sample} contains 29 Li-rich giant stars, among which we obtained asteroseismic evolutionary phases for 18 stars.

%----------------------------------------------------------------------------------------
%   Observation \& Data Reduction
%----------------------------------------------------------------------------------------
\vspace{10pt}
\noindent\textbf{Observation \& Data Reduction} 

For our {\it low-resolution sample}, they have been observed during the LAMOST-Kepler project and the regular survey. The corresponding data are reduced and released by pipelines of LAMOST\upcite{Luo2015} and Kepler\upcite{Jenkins2010}. For the high resolution sample, the targets were observed with five telescopes, including the 8.2-meter Subaru telescope (Japan) at Mauna Kea Observatory, Hawaii, 3.5-meter telescope at Apache Point Observatory (APO), New Mexico, the 2.4-meter Automated Planet Finder (APF) telescope at Lick Observatory, California, 2.4-meter and 1.8-meter telescope at Lijiang Observatory, Yunan Province. The observation information are listed in Supplementary Table~\ref{sup_tab_1}. For the spectra observed with Subaru, we use an {\it iraf} standard package for data reduction, while for the spectra observed by other telescopes, we use a package based on Interactive Data Language ({\it IDL}) environment to reduce the data. Both reductions follow the same procedures, including bias and flat subtracting, order tracing, wavelength calibration, instrumental response correcting, background scatter subtracting and cosmic rays removing.

%----------------------------------------------------------------------------------------
%   Stellar Parameters and Elemental Abundances of the High-resolution Sample
%----------------------------------------------------------------------------------------
\vspace{10pt}
\noindent\textbf{Stellar Parameters, Li Abundances, and Error Estimation of the High-resolution Sample}

Stars studied with the high-resolution spectroscopy were selected prior to any asteroseismic analysis. For the stars with the high-resolution spectra, we use the spectroscopic method to derive their stellar parameters by requiring the ionization and excitation equilibriums for \ion{Fe}{1} and \ion{Fe}{2} lines. The Fe line list used in this work is as same as that used in our previous work\upcite{Yan2018}, which is a combination of three Fe line lists\upcite{Takeda2002, Mashonkina2011, Carlberg2012}. The atomic line data of Fe have been calibrated with solar spectrum\upcite{Kurucz1984}. We use unblended lines with moderate strength ($20 - 110$\,m\AA) and excitation energy (\Eexc) greater than $2.0$\,eV\upcite{Sitnova2015} for each star in our sample. The effective temperature (\Teff) is determined by excitation equilibrium of \ion{Fe}{1} lines. The micro-turbulence velocity (\Vt) is constrained by requiring that the Fe abundances derived from individual \ion{Fe}{1} lines are independent to their equivalent widths. The surface gravity (\logg) is obtained by minimizing the Fe abundances derived from \ion{Fe}{1} and \ion{Fe}{2} lines, and the metallicity ([Fe/H]) is averaged from iron abundances derived from the \ion{Fe}{2} lines.

To estimate the random errors of the stellar parameters, we compare our result to those derived from an independent study or method. We find 22 stars are in common between our {\it high-resolution sample} and ASPCAP\upcite{Garcia2016} DR15. We compare our \Teff\ with \Teff\ from ASPCAP in Supplementary Fig.~\ref{sup_fig_1} (top panel). We find a good agreement within the two data sets. The systematic error can be ignored and the standard deviation is $54$ K. Thus we adopted this value as our estimated random error of \Teff\ in the {\it high-resolution sample}.

We note that all the stars in our {\it high-resolution sample} have {\it Gaia} parallaxes, which can be used as an independent way of determining the surface gravity. For each star, we first calculate its bolometric magnitude $M_{\rm bol}$ from $\displaystyle M_{\rm bol}=V_{\rm mag}-5\log(d)+5-A_{\rm V}+BC$, where $V_{\rm mag}$ is the magnitude of the star in V band, $d$ is the distance estimated from {\it Gaia} parallax by Bailer-Jones et al. (2018)\upcite{Bailer-Jones2018} applying a weak distance prior to the Galaxy model, $A_{\rm V}$ is obtained from the Galactic extinction map provided by Schlafly \& Finkbeiner in 2011\upcite{Schlafly2011}, and the bolometric correction BC is calculated following the method of Alonso \emph{et al.}\upcite{Alonso1999}. The surface gravity form {\it Gaia} parallax then can be calculated by $\displaystyle \log g_{\rm gaia} = \log g_{\odot}+\log ({M}/{M_{\odot}})+4\log({T_{\rm eff}}/{T_{\rm eff\odot}})+0.4(M_{\rm bol}-M_{\rm bol\odot})$, where the solar values are adopted as $\log g_{\odot}=4.44$, $T_{\rm eff\odot}=5777$\,K, and $M_{\rm bol\odot} = 4.74$\,mag. We find our surface gravities show a good consistency with those derived from {\it Gaia} parallaxes, as shown in Supplementary Fig.~\ref{sup_fig_1} (bottom panel), with a scatter of $\sim 0.13$ {\it dex}. Similarly, we adopt this value as our estimated error of the surface gravities.

We also compared the metallicities of our sample to the ASPCAP results, and we find a systematic difference of 0.13 {\it dex} with a scatter of 0.14 {\it dex}. The systematic difference is most likely to be caused by the differences of the adopted surface gravities. Since we did not find any evident systematic difference between our \logg and $\log g_{\rm Gaia}$, we thus only use the scatter on [Fe/H] as our estimated error for metallicity, which is $\sim 0.14$ {\it dex}.

The Li abundances in the high-resolution spectra are derived from a spectral synthesis method. The synthesized line profiles are calculated based on from the {\it MARCS}\upcite{Gustafsson2008} model atmospheres. The equations of coupled radiative transfer and statistical equilibrium for NLTE calculations are solved by a revised {\it DETAIL} program using accelerated lambda iteration method (for more details, see Mashonkina \emph{et al.} 2011\upcite{Mashonkina2011}). Two lines are used for deriving the Li abundance, namely the resonance line at $6707.8$\,\AA\ and the subordinate line at $6103.6$\,\AA. The atomic model used for NLTE analysis is presented by Shi et al. (2007)\upcite{Shi2007}. The adopted Li abundance is an average result derived from these two lines, and the errors are estimated from the abundance differences derived from these two lines. We present the results in Supplementary Table~\ref{sup_tab_2}.

%----------------------------------------------------------------------------------------
%   Stellar Parameters and Li abundances for the Low-resolution Sample
%----------------------------------------------------------------------------------------
\vspace{10pt}
\noindent\textbf{Stellar Parameters, Li Abundances, and Error Estimation for the Low-resolution Sample} 

For our {\it low-resolution sample}, we adopt the stellar parameters (effective temperature, surface gravity, and metallicity) derived from LAMOST pipeline\upcite{Luo2015} of DR7. The Li abundances are derived using a template matching method. The templates are synthesized using the SPECTRUM code with Kurucz ODFNEW model atmospheres\upcite{Castelli2003}. The standard solar composition is adopted from Grevesse \& Sauval (1998)\upcite{Grevesse1998}. The synthesized templates were convolved by a set of Gaussian profiles to match the broadening (dominated by instrument) of LAMOST spectra. We obtained a set of grid in the stellar parameters space with steps of $100$K, $0.25$ {\it dex}, and $0.20$ {\it dex} for effective temperature, surface gravities, and metallicities, respectively. For the step of Li abundance, we set a 0.10 {\it dex} interval in the range of $-3.0<$ [Li/Fe] $<6.9$.

The Li abundances are derived from \ion{Li}{1} resonance line at $6707.8$\,\AA. We first generate a set of synthesized spectra based on the fixed stellar parameters and the grid of Li abundances, then we calculate the chi-square of each template to the observed spectra. Since the grid of Li abundance is a set of discrete values, the chi-square obtained is also a discrete array. We fit a curve to the discrete chi-square array and find its minimum. Each minimum chi-square has two adjacent points in the chi-square array. The Li abundance is then interpolated based on these two values in Li abundance grid. We plot some of our matching results in Supplementary Fig.~\ref{sup_fig_2}. Stars shown in this figure are selected based on their Li abundances, which is from $\sim 1.5$ {\it dex} to $\sim 3.1$ {\it dex}. Finally, we eliminate the spurious results by 1) an automatic self-inspection and 2) eye-inspection (see Gao et al. 2019\upcite{Gao2019} for details). 

The random errors of Li abundance in our low-resolution spectra are estimated in two ways. The first one is the comparison with the high-resolution results as we have 26 stars in common with our {\it high-resolution sample}. By using the stellar parameters obtained from the high-resolution spectra, we calculated the Li abundances for the common sources in the {\it low-resolution sample}, and compare them with the LTE Li abundances obtained from the high-resolution spectra (Supplementary Fig.~\ref{sup_fig_3}), and obtain a standard deviation of $\pm 0.24$ {\it dex}. Another way to estimate the error is to calculate the Li abundance for the targets that have more than one observation in LAMOST survey. For example, if a target has three exposures, we treat them as three different stars and match our templates to each of the spectrum, then the standard deviation of the three results is marked as the error of the star. We estimated the error of the sample by averaging all the standard deviations marked in the previous process. The error estimated in such way is $\pm 0.21$ {\it dex}. The uncertainties of the stellar parameters also result in the uncertainties of the Li abundances. To evaluate this uncertainty, we randomly chose one star with the typical stellar parameters and Li abundance in our sample. We change the stellar parameters within a typical error range, namely $\pm$ 100\,K, $\pm$ 0.2\,dex, $\pm$ 0.2\,dex for the \Teff, \logg, and [Fe/H], respectively. The final Li abundance varies with the change of stellar parameters. In general, uncertainty on \Teff\ affects the Li abundance most. An uncertainty of 100K would result in about 0.15 dex uncertainty on the Li abundance. We present the detailed results in Supplementary Table~\ref{sup_tab_3}.

%----------------------------------------------------------------------------------------
%   The Asteroseismic Analysis \& Evolutionary Phase
%----------------------------------------------------------------------------------------
\vspace{10pt}
\noindent\textbf{The Asteroseismic Analysis \& Evolutionary Phase}

By building a convolutional neural network model to the power spectra of red giants, Hon et al. (2017) successfully obtained the classifications of evolutionary phases\upcite{Hon2017} to a large sample of red giants, and the asteroseismic parameters \dnu, \numax, masses and radii can be obtained\upcite{Yu2018}. 

We visually examined the spacings between consecutive $l = 1$ mixed modes to a fraction of our sample stars, and found a good reliability of their results. Thus we adopted their classifications and obtained the evolutionary phases as the initial results. We found 134 Li-rich stars in our sample have such asteroseismic parameters. Then, based on the asteroseismic patterns, we calculated the period spacings of {\it g}-mode to the stars with high signal-to-noise ratios by matching observed spectra with templates which were constructed using the asymptotic theory for mixed modes\upcite{Unno1989, Shibahashi1979, Mosser2015}. We classify these stars using the classic \dpi\ versus \dnu\ diagram\upcite{Bedding2011}. We find the classification by this method is consistent with our initial results except for two stars with masses closed to 2.4 M$_{\odot}$. We adopted the \dpi $=150 s$ as the separation criterion.

We also derived the masses and radii for the lithium-rich giants using the scaling relations with \dnu\ and \numax\upcite{Brown1991, Kjeldsen1995}. We followed Sharma et al. (2016)\upcite{Sharma2016} to correct the uncertainties of scaling relations. The correction to the relation with \dnu\ is obtained from grid models, and varies with metallicity, mass and age. 

We derived the evolutionary phases for the stars in the {\it high-resolution sample} by the combination of asteroseismic analysis and the H-R diagram. The evolutionary tracks are obtained from {\it PARSEC} tracks\upcite{Bressan2012}. Stars are divided into eight groups based on their metallicities for the corresponding tracks. The luminosities of stars were derived from the bolometric magnitude using  $\displaystyle \log L = -0.4 \times (M_{\rm bol}-M_{\rm \odot})$. The uncertainty of luminosity is mainly from the distance and extinction that used to calculate $M_{\rm bol}$. The errors of the distance are from Bailer-Jones et al. (2018)\upcite{Bailer-Jones2018}. The stars in our {\it high-resolution sample} are not very distant in general, the relative error is below 15\% for the distance. The errors of the extinction is hard to evaluate. We estimate this error as 0.05 for the color excess E(B-V), which is larger than that presented by Schlafly \& Finkbeiner 2011\upcite{Schlafly2011}. The error of luminosity is calculated based on the error propagation equation. The typical error for the luminosity is about 0.1 dex on logarithmic scale.

In Supplementary Fig.~\ref{sup_fig_4}, we place our Li-rich giants in the {\it high-resolution sample} on the H-R diagram with the spectroscopic stellar parameters. From the asteroseismology, we obtained the evolutionary stage of 18 Li-rich giants. The other 11 Li-rich giants without asteroseismic data are classified based on the following criterions: 1) For a star in the {\it overlapping region}, if its $A_{\rm Li} \ge 2.6$ {\it dex}, then it is classified as an RC star; otherwise, it is classified based on its location to the closest track. 2) For a star not in the {\it overlapping region}, it is classified based on its location to the closest track. Stars with low \logg\ (around 1.5 or less) are highly evolved RGB or AGB stars.

%----------------------------------------------------------------------------------------
%   The Statistical Test to the Sample
%----------------------------------------------------------------------------------------
\vspace{10pt}
\noindent\textbf{The Statistical Test to the Sample}

The distribution of Li-rich RGB stars is fitted with the NLS function from R language. Both exponential and linear fitting are tested. For the exponential fitting, we obtained the best fit as {$\displaystyle y=e^{(-3.55x_{i}+6.39)}\times 100\,\%$}, with a residual standard error of 0.047. For the linear fitting, the best fit for the distribution is {$ \displaystyle y=(-1.82x_{i}+4.21)\times 100\,\%$}, with a residual standard error of 0.149. We adopted the exponential fitting for the distribution.

It is also very important to test if the signatures found in the stars with the asteroseismology data (134 stars) could represent the signatures of the whole sample (455 stars). We used the Kolmogorov-Smirnov(KS) test to do this. The KS test is a classical method to examine whether a set of observations are from some completely specified continuous distribution \upcite{Lilliefors1969}. Barr \& Davidson 1973\upcite{Barr1973} discussed the Kolmogorov-Smirnov `goodness-of-fit' test for its use with censored or truncated samples. For our sample, the distribution of the stars with asteroseismology data (134 stars) is defined as {\it d1}, and the distribution of the whole sample (455 stars) is defined as {\it d2}. We found the probability that {\it d1} and {\it d2} have the same distribution is 0.94, and the maximum difference between them is 0.05. Thus, we consider the distribution of 134 stars have the same distribution as the whole sample.

%----------------------------------------------------------------------------------------
%   The Calculation of HeWD+RGB Merger Model
%----------------------------------------------------------------------------------------
\vspace{10pt}
\noindent\textbf{The Calculation of HeWD+RGB Merger Model} 

We obtain the information about potential merger progenitors by binary star population synthesis. We use a rapid binary evolution code (BSE)\upcite{Hurley2000, Hurley2002} to evolve $10^7$ pairs of zero-age main-sequence stars for 14 Gyr. Then we record the properties of HeWD+RGB binaries at the onset of the common envelope (CE) phase. The information of such pre-CE binaries will be used to set the grid of parameters for the calculations of post-mergers. The settings in the BSE code in this work are chosen to be similar to the previous studies\upcite{Izzard2007, Zhang2014, Zhang2017}.

We use the stellar evolution code of Modules for Experiments in Stellar Astrophysics (MESA) v8118\upcite{Paxton2011,Paxton2013,Paxton2015} to examine the feature of post-merger including enrichment of the elements. We use a series of separate accretion steps to simulate a merger with a 1D stellar evolution code, which has previously been used successfully to represent some observations of merger remnants.\upcite{Zhang2013, Zhang2014, Zhang2017} In the following subsequent of post-merger evolution, we adopted parameters similar to MESA isochrones and stellar tracks (MIST) project for normal stars\upcite{Dotter2016, Choi2016}. In our models, mixing is by convection in the convective regions and atomic diffusion in the radiative areas\upcite{Thoul1994}. Diffusion includes the processes of gravitational settling, thermal diffusion, and concentration diffusion. We also considered semi-convective and thermohaline mixing as in MIST.

According to the CE merging process, the remnant contains a hybrid core with a hot helium shell ($> 10^8$\,K) surrounded by a hydrogen envelope. At the early stage of the merger process, $^3$He from the hydrogen envelope is mixed with $^4$He in hot helium shell and produces the fresh $^7$Li by the $^3$He$(\alpha; \gamma)$$^7$Be$(e-; \nu)$$^7$Li reaction. Then, the convection zone will shrink away from the hot shell and back to a region where the temperature is less than $2.5\times10^6$\,K, leaving some newborn $^7$Li to survive in the surface. Hence, we obtain some mergers with lithium enrichment. 

We obtain the distribution of Li-rich giants by combining both results of binary star population synthesis and evolutionary tracks of post-mergers. Four metallicities are included in our calculation, i.e. 0.03, 0.02, 0.01, and 0.004. By our calculation of 107 binary systems, there are 3931, 3233, 2707 and 3093 pairs undergo HeWD+RGB mergers with enriched lithium surfaces for metallicities Z = 0.03, 0.02,0.01 and 0.004, respectively. The masses of Li-rich giants are in a range from 0.8 to 1.8 $M_{\odot}$ with a peak at 1.1-1.2 $M_{\odot}$.

%----------------------------------------------------------------------------------------
%additional references
%----------------------------------------------------------------------------------------

\vspace{10pt}
\noindent \textbf{Data availability} {The data that support the plots within this paper and other findings of this study are available from the corresponding authors upon reasonable request. The data for Figs.\ref{fig_1}-\ref{fig_3} are provided as the source data files. The LAMOST DR7 data are available for registered users from \url{http://dr7.lamost.org/}. The stellar evolution tracks and isochrones data of PARSEC are available from \url{https://people.sissa.it/~sbressan/CAF09_V1.2S_M36_LT/}.}

\vspace{10pt}
\noindent \textbf{Code availability} {The code that support the plots within this paper and other findings of this study are available from the corresponding authors upon reasonable request. The stellar evolution code MESA used to compute the HeWD-RGB merger is available from \url{http://mesa.sourceforge.net}. The code SPECTRUM used to compute the stellar templates for deriving the Li abundance for the low-resolution sample is available from \url{http://www.appstate.edu/~grayro/spectrum/spectrum.html}. The codes used to calculate the luminosity and surface gravity from the Gaia parallax are available from \url{https://github.com/YutaoZhou/2020Na_codes/}. The codes involved in the plots are based on the astrolib, Coyote Library, Matplotlib, Pandas, Numby and Astropy. }

%----------------------------------------------------------------------------------------
%   Supplementary Information
%----------------------------------------------------------------------------------------
\setcounter{figure}{0}
\setcounter{table}{0}
\captionsetup[figure]{labelfont={bf},name={Supplementary Fig.},labelsep=period}
\captionsetup[table]{labelfont={bf},name={Supplementary Table},labelsep=period}

\clearpage
%\section*{Supplementary Information}
%----------------------------------------------------------------------------------------
%   Supplementary Figures and Tables
%----------------------------------------------------------------------------------------

\clearpage
%%%%%%%%%%%%%%%%%%%%%%%%%%%%%%%%%%%%  Supplementary Figure 1  %%%%%%%%%%%%%%%%%%%%%%%%%%%%%%%%%%%% 
\begin{figure*}[!t]
\begin{center}
\includegraphics[angle=0, width=10cm]{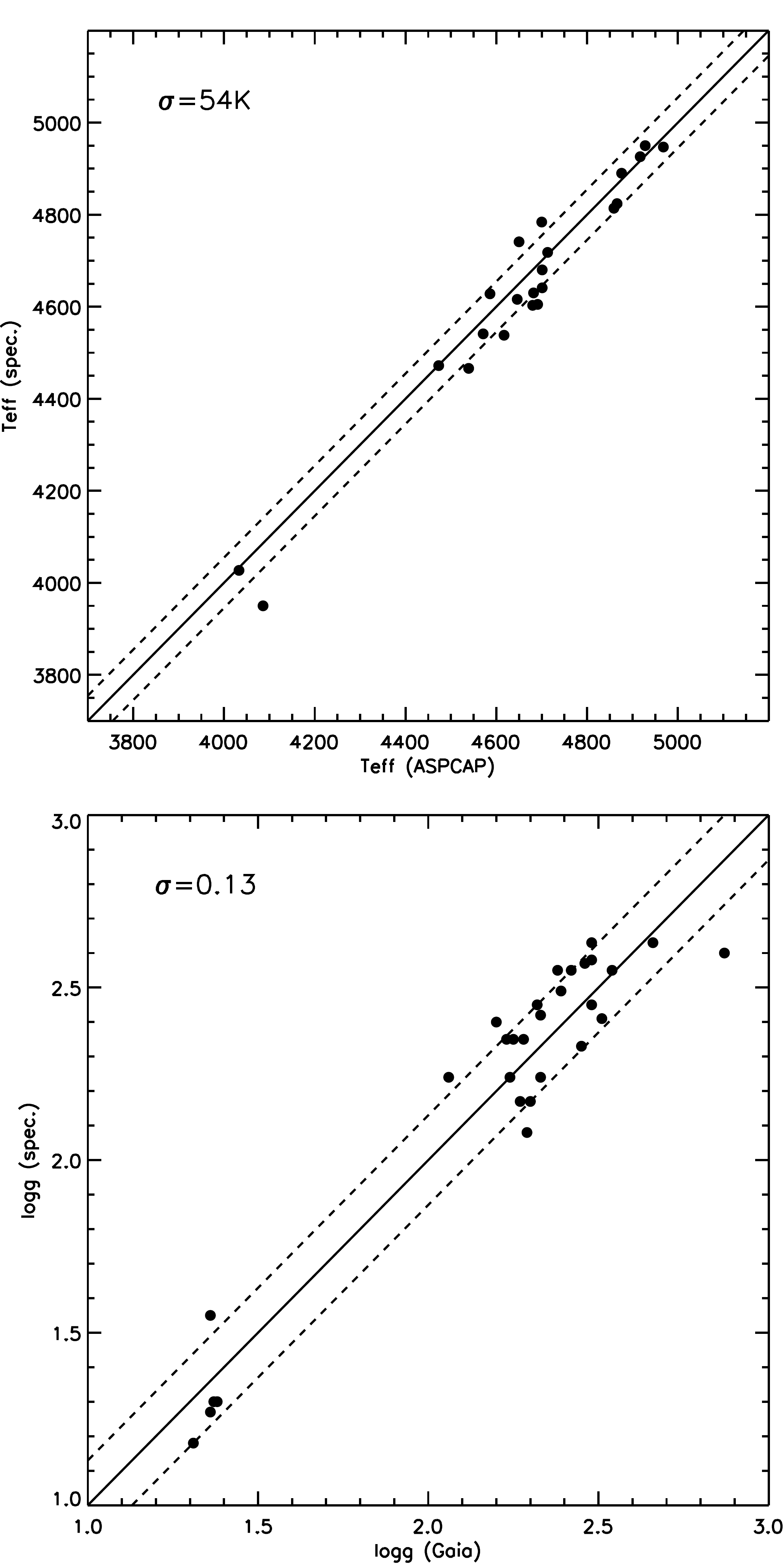}\\
\end{center}
\caption{\textbf{The comparison of the stellar parameters between spectroscopic method and other methods.} The \Teff\ used for comparison is obtained from ASPCAP, and the \logg\ is calculated from {\it Gaia} parallaxes. The solid line represents a 1:1 ratio, and the dashed line marked the standard deviation in each panel. The number of stars for compare and the standard deviation is labeled in each panel.}\label{sup_fig_1}
\end{figure*}
%%%%%%%%%%%%%%%%%%%%%%%%%%%%%%%%%%%%  Supplementary Figure 1  %%%%%%%%%%%%%%%%%%%%%%%%%%%%%%%%%%%% 

\clearpage
%%%%%%%%%%%%%%%%%%%%%%%%%%%%%%%%%%%%  Supplementary Figure 2  %%%%%%%%%%%%%%%%%%%%%%%%%%%%%%%%%%%% 
\begin{figure*}[!t]
\begin{center}
\includegraphics[angle=0, width=15cm]{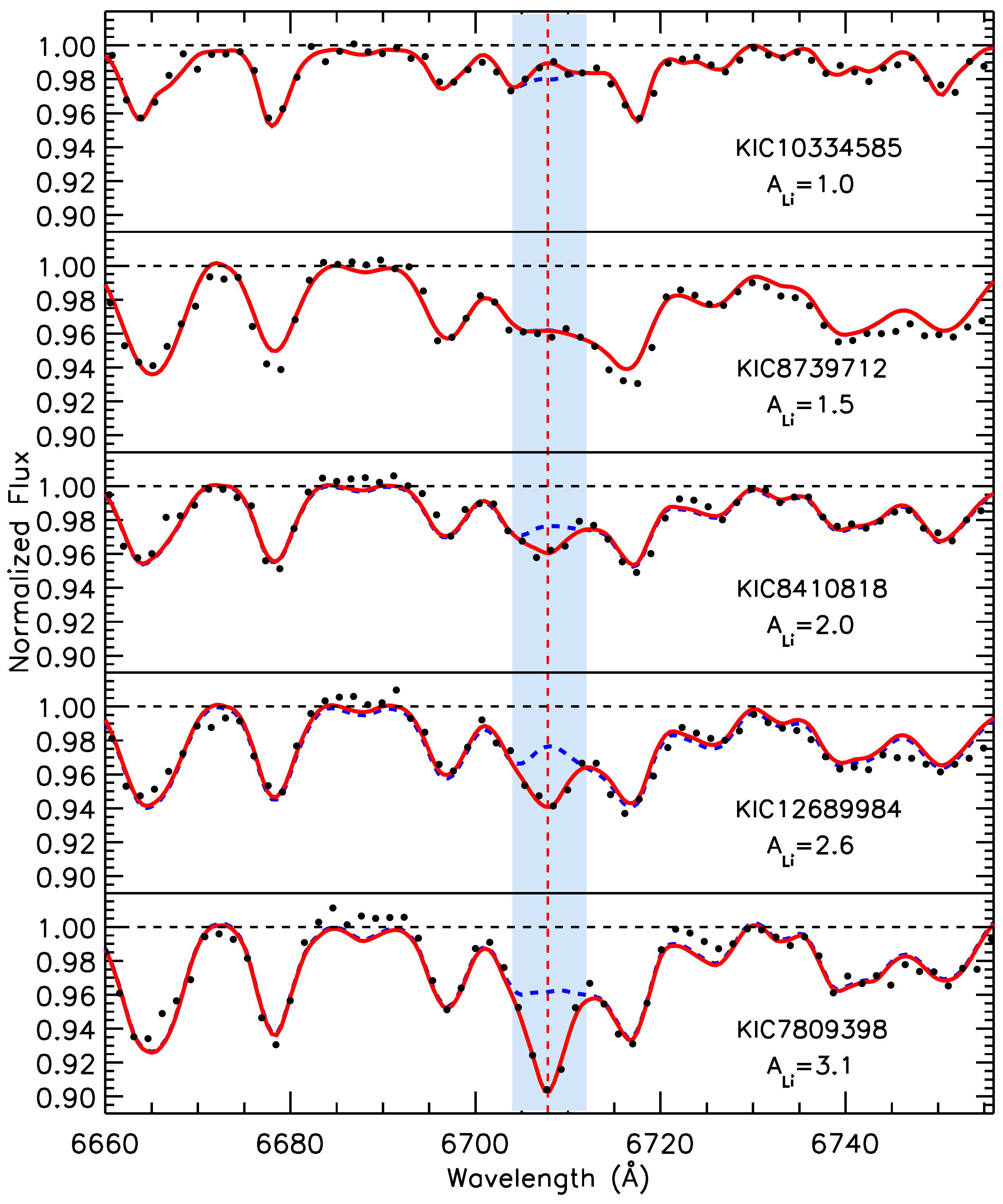}\\
\end{center}
\caption{\textbf{The LAMOST spectra and their best template matching.} The black dots represent the observed low-resolution spectra by LAMOST, and the red solid lines are the best-fitted templates. The templates are calculated with stellar parameters from LAMOST DR7 and the fitted Li abundance. The color of light blue indicates the region that is used to calculate the chi-square during the fitting. The Li abundances and star ID are given in each panel. Note the vertical axis is the normalized flux and has been zoomed into the range of $0.90-1.02$.}\label{sup_fig_2}
\end{figure*}
%%%%%%%%%%%%%%%%%%%%%%%%%%%%%%%%%%%%  Supplementary Figure 2  %%%%%%%%%%%%%%%%%%%%%%%%%%%%%%%%%%%% 

\clearpage
%%%%%%%%%%%%%%%%%%%%%%%%%%%%%%%%%%%%  Supplementary Figure 3  %%%%%%%%%%%%%%%%%%%%%%%%%%%%%%%%%%%% 
\begin{figure*}
\begin{center}
\includegraphics[angle=0, width=\hsize]{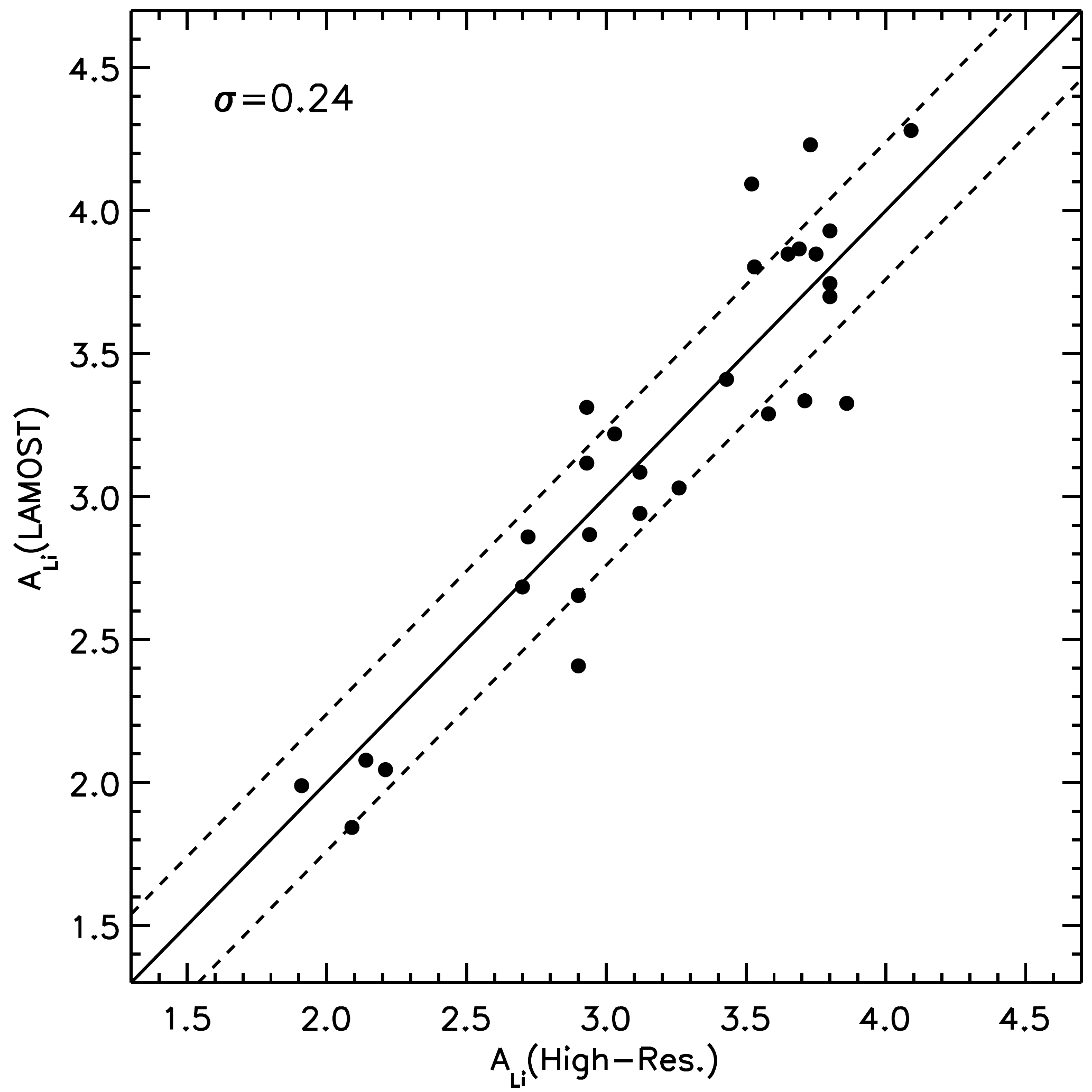}\\
\end{center}
\caption{\textbf{The comparison between the Li abundances derived from the low- and high-resolution data.} The Li abundances used here are derived from 6708\,\AA\ line. Symbols and labels are similar to Supplementary Fig.~\ref{sup_fig_1}.}\label{sup_fig_3}
\end{figure*}
%%%%%%%%%%%%%%%%%%%%%%%%%%%%%%%%%%%%  Supplementary Figure 3  %%%%%%%%%%%%%%%%%%%%%%%%%%%%%%%%%%%% 

\clearpage
%%%%%%%%%%%%%%%%%%%%%%%%%%%%%%%%%%%%  Supplementary Figure 4  %%%%%%%%%%%%%%%%%%%%%%%%%%%%%%%%%%%% 
\begin{figure*}[!t]
\begin{center}
\includegraphics[angle=0, width=\hsize]{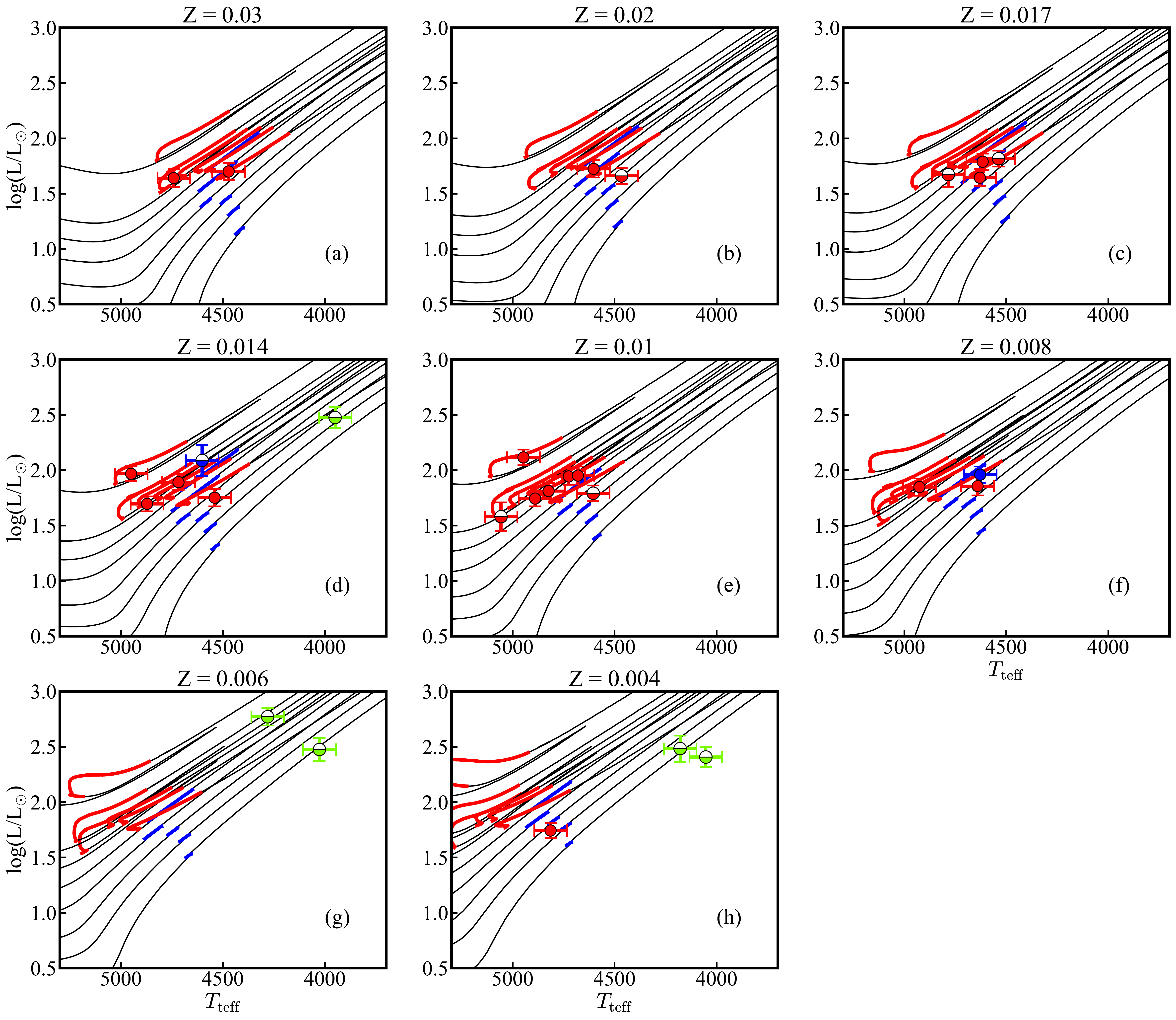}\\
\end{center}
\caption{\textbf{The locations of 29 Li-rich giants on the H-R diagram.} The tracks are plotted from 0.8 M$_{\odot}$ to 2.0 M$_{\odot}$ with interval of $0.2$ M$_{\odot}$, plus a track of 2.6 M$_{\odot}$ (the top one). Stars are divided into eight groups based on their metallicity for corresponding tracks. The symbols are as same as that in Fig.~2.}\label{sup_fig_4}
\end{figure*}
%%%%%%%%%%%%%%%%%%%%%%%%%%%%%%%%%%%%  Supplementary Figure 4  %%%%%%%%%%%%%%%%%%%%%%%%%%%%%%%%%%%% 

\clearpage
\begin{table*}[!t]
\begin{center}
\caption{\textbf{The observation information of Li-rich giants in the high-resolution sample.}}\label{sup_tab_1}
\begin{tabular}{rrcccccr}
\hline\hline\noalign{\smallskip}
         ID1  &       KIC  &  Vmag   &  Obs.Date   &  Telescope     & Instrument &  Res.                      &  SNR  \\  
              &            &  mag    &             &                &            &  $\lambda / \Delta\lambda$ &       \\
\noalign{\smallskip}
\hline\noalign{\smallskip}
  J1849+4840  &  11068543  & 11.530 &  2017-08-05  &  Subaru        & HDS        &  $\sim 45,000$            &  104 \\  
  J1854+4316  &   7587353  & 11.739 &  2015-05-29  &  APF           & ""         &  $\sim 80,000$            &  87 \\  
  J1859+4507  &   8869656  &  9.368 &  2017-10-06  &  Lijiang-2.4m  & HiRES      &  $\sim 30,000$            &  88  \\  
  J1901+4944  &  11651091  & 11.393 &  2017-08-05  &  Subaru        & HDS        &  $\sim 45,000$            &  85  \\  
  J1905+4834  &  11019752  & 11.076 &  2017-08-05  &  Subaru        & HDS        &  $\sim 45,000$            &  92  \\  
  J1907+5000  &  11805390  &  9.804 &  2016-10-09  &  APF           & ""         &  $\sim 80,000$            &  97 \\  
  J1917+5145  &  12645107  & 11.442 &  2017-08-04  &  Subaru        & HDS        &  $\sim 45,000$            &  77  \\  
  J1920+5203  &  12784683  & 11.462 &  2016-10-11  &  APF           & ""         &  $\sim 80,000$            &  73  \\  
  J1928+3741  &   2305930  & 11.111 &  2015-05-20  &  APO-3.5m      & ARCES      &  $\sim 32,000$            &  123 \\  
  J1929+4423  &   8366758  & 12.500 &  2017-08-04  &  Subaru        & HDS        &  $\sim 45,000$            &  104 \\  
  J1930+4942  &  11663387  & 12.591 &  2016-11-16  &  Subaru        & HDS        &  $\sim 45,000$            &  82  \\  
  J1931+4521  &   9024667  & 12.286 &  2017-08-04  &  Subaru        & HDS        &  $\sim 45,000$            &  76  \\  
  J1931+4639  &   9833651  & 12.519 &  2016-11-17  &  Subaru        & HDS        &  $\sim 45,000$            &  94  \\  
  J1934+4108  &   5881715  & 11.781 &  2016-10-19  &  APF           & ""         &  $\sim 80,000$            &  80 \\  
  J1934+3858  &   3858850  & 12.444 &  2016-11-16  &  Subaru        & HDS        &  $\sim 45,000$            &  67  \\  
  J1936+4938  &  11615224  & 11.159 &  2015-09-04  &  APO-3.5m      & ARCES      &  $\sim 32,000$            &  118 \\  
  J1937+4339  &   7898227  & 12.915 &  2016-11-16  &  Subaru        & HDS        &  $\sim 45,000$            &  68  \\  
  J1938+4006  &   5021453  & 11.381 &  2018-09-16  &  APO-3.5m      & ARCES      &  $\sim 32,000$            &  146 \\  
  J1938+4338  &   7899597  & 13.610 &  2016-11-17  &  Subaru        & HDS        &  $\sim 45,000$            &  81  \\  
  J1939+4614  &   9596106  & 11.664 &  2017-08-05  &  Subaru        & HDS        &  $\sim 45,000$            &  75  \\  
  J1945+4337  &   7905528  & 11.924 &  2015-06-23  &  APF           & ""         &  $\sim 80,000$            &  90 \\  
  J1949+4750  &  10616012  & 11.454 &  2015-06-11  &  APO-3.5m      & ARCES      &  $\sim 32,000$            &  126 \\  
  J1951+4618  &   9665729  &  9.302 &  2017-09-25  &  Lijiang-1.8m  & HRS        &  $\sim 37,000$            &  114 \\  
  J1951+4647  &   9907856  & 12.451 &  2017-08-04  &  Subaru        & HDS        &  $\sim 45,000$            &  89  \\  
  J1956+4116  &   5989157  & 12.278 &  2016-11-17  &  Subaru        & HDS        &  $\sim 45,000$            &  132 \\  
  J1957+4550  &   9364778  & 11.204 &  2017-08-04  &  Subaru        & HDS        &  $\sim 45,000$            &  96  \\  
  J0331+1933  & 210778970  & 13.442 &  2016-11-17  &  Subaru        & HDS        &  $\sim 45,000$            &  93  \\  
  J0828+2244  & 212136170  & 13.318 &  2016-11-16  &  Subaru        & HDS        &  $\sim 45,000$            &  39  \\  
  J1119-0104  & 201392458  & 13.439 &  2017-02-19  &  Subaru        & HDS        &  $\sim 45,000$            &  45  \\ 
\noalign{\smallskip} \hline
\end{tabular}
\end{center}
\end{table*}

\clearpage
\begin{table*}[!t]
\begin{center}
\caption{\textbf{The spectroscopic results of Li-rich giants in the high-resolution sample.} The evolutionary phase are indicated in the seventh column, with `1' represents RGB stars, `2' represents RC stars, and `3' represents highly evolved RGB or AGB stars for which asteroseismology data is not available. The `star'(*) indicates that the phase is derived from asteroseismic analysis. The phase without * is derived using the criterion described in this paper.}\label{sup_tab_2}
\begin{tabular}{rrcccclr}
\hline\hline\noalign{\smallskip}
          id  &       KIC &  \Teff & \logg & $[$Fe/H$]$ &   \Vt &   phase &  A(Li)$_{\rm NLTE}$  \\
              &           &      K &       &            &  \kms &         &                      \\
\noalign{\smallskip}
\hline\noalign{\smallskip}
  J1849+4840  &    11068543 &  3950 &  1.30 &  -0.15  &  1.15  &       3   &    2.83 $\pm$ 0.05  \\
  J1854+4316  &     7587353 &  4616 &  2.24 &   0.00  &  1.46  &       2*  &    3.48 $\pm$ 0.10  \\
  J1859+4507  &     8869656 &  4872 &  2.55 &  -0.10  &  0.97  &       2*  &    3.61 $\pm$ 0.09  \\
  J1901+4944  &    11651091 &  4466 &  2.55 &   0.10  &  1.30  &       2   &    2.65 $\pm$ 0.09  \\
  J1905+4834  &    11019752 &  4605 &  2.42 &  -0.21  &  1.40  &       2   &    2.81 $\pm$ 0.08  \\
  J1907+5000  &    11805390 &  4950 &  2.58 &  -0.06  &  1.84  &       2*  &    2.24 $\pm$ 0.10  \\
  J1917+5145  &    12645107 &  4824 &  2.55 &  -0.24  &  1.15  &       2*  &    3.30 $\pm$ 0.09  \\
  J1920+5203  &    12784683 &  4890 &  2.63 &  -0.25  &  1.28  &       2*  &    2.79 $\pm$ 0.08  \\
  J1928+3741  &     2305930 &  4814 &  2.35 &  -0.59  &  1.37  &       2*  &    3.90 $\pm$ 0.03  \\
  J1929+4423  &     8366758 &  4741 &  2.17 &   0.20  &  1.49  &       2*  &    3.24 $\pm$ 0.10  \\
  J1930+4942  &    11663387 &  4630 &  2.45 &   0.00  &  1.25  &       2*  &    3.43 $\pm$ 0.04  \\
  J1931+4521  &     9024667 &  4541 &  2.35 &  -0.06  &  1.31  &       2*  &    2.79 $\pm$ 0.10  \\
  J1931+4639  &     9833651 &  4603 &  2.41 &   0.11  &  0.97  &       2*  &    3.40 $\pm$ 0.09  \\
  J1934+4108  &     5881715 &  4725 &  2.33 &  -0.18  &  1.28  &       2*  &    3.38 $\pm$ 0.11  \\
  J1934+3858  &     3858850 &  4472 &  2.17 &   0.26  &  1.45  &       2*  &    2.95 $\pm$ 0.09  \\
  J1936+4938  &    11615224 &  4680 &  2.24 &  -0.22  &  1.30  &       2*  &    2.84 $\pm$ 0.02  \\
  J1937+4339  &     7898227 &  4179 &  1.30 &  -0.76  &  1.70  &       3   &    2.81 $\pm$ 0.13  \\
  J1938+4006  &     5021453 &  4718 &  2.40 &  -0.10  &  1.20  &       2*  &    3.62 $\pm$ 0.08  \\
  J1938+4338  &     7899597 &  4641 &  2.45 &  -0.40  &  1.60  &       2*  &    3.39 $\pm$ 0.06  \\
  J1939+4614  &     9596106 &  4628 &  2.24 &  -0.31  &  1.52  &       1*  &    2.18 $\pm$ 0.06  \\
  J1945+4337  &     7905528 &  4027 &  1.18 &  -0.49  &  1.38  &       3   &    3.49 $\pm$ 0.01  \\
  J1949+4750  &    10616012 &  4538 &  2.35 &   0.02  &  1.30  &       2   &    3.48 $\pm$ 0.08  \\
  J1951+4618  &     9665729 &  4281 &  1.55 &  -0.46  &  1.95  &       3   &    2.13 $\pm$ 0.03  \\
  J1951+4647  &     9907856 &  4926 &  2.49 &  -0.30  &  1.25  &       2*  &    2.51 $\pm$ 0.05  \\
  J1956+4116  &     5989157 &  4053 &  1.27 &  -0.73  &  1.80  &       3   &    3.05 $\pm$ 0.02  \\
  J1957+4550  &     9364778 &  4947 &  2.57 &  -0.16  &  1.44  &       2*  &    3.43 $\pm$ 0.03  \\
  J0331+1933  &   210778970 &  5057 &  2.60 &  -0.22  &  1.25  &       2   &    3.08 $\pm$ 0.10  \\
  J0828+2244  &   212136170 &  4602 &  2.08 &  -0.14  &  0.70  &       1   &    1.92 $\pm$ 0.08  \\
  J1119-0104  &   201392458 &  4784 &  2.63 &   0.02  &  1.10  &       2   &    2.95 $\pm$ 0.09  \\
\noalign{\smallskip} \hline
\end{tabular}
\end{center}
\end{table*}

\clearpage
\begin{table*}[!t]
\begin{center}
\caption{\textbf{The uncertainties of the Li abundance caused by the uncertainties of the stellar parameters}. The stellar parameters (namely \Teff, \logg, and [Fe/H]) of the two stars are shown in the bracket following their IDs.}\label{sup_tab_3}
\begin{tabular}{cccccccc}
\hline\hline\noalign{\smallskip}
                             & \Teff\ (K) & \Teff\ (K) & \logg     & \logg     & $[$Fe/H$]$ & $[$Fe/H$]$ & ID                        \\
                             & +100      & -100      & +0.2      & -0.2      & +0.2       & -0.2       &                           \\
\noalign{\smallskip}
\hline\noalign{\smallskip}
 uncertainty of A$_{\rm Li}$ & +0.15     & -0.17     & +0.00     & -0.01     & +0.00      & -0.01      &  Star 1 (4699, 2.42, 0.0)   \\
                             & +0.16     & -0.18     & +0.01     & -0.01     & +0.00      & -0.00      &  Star 2 (4589, 2.44, -0.4)  \\
\noalign{\smallskip} \hline

\end{tabular}
\end{center}
\end{table*}

\end{document}